\documentclass[twocolumn]{jpsj2} 
\usepackage{amsmath}
\usepackage{amssymb}
\usepackage{graphicx}
\usepackage{graphics}
\usepackage{overcite}

\title{
 Pairing competition in a quasi-one-dimensional model 
 of organic superconductors (TMTSF)$_{2}X$ in magnetic field
}

\author{
 Hirohito \textsc{Aizawa}$^{1}$, 
 Kazuhiko \textsc{Kuroki}$^{1}$, 
 and Yukio \textsc{Tanaka}$^{2}$
}

\inst{
$^{1}$Department of Applied Physics and Chemistry, 
 The University of Electro-Communications, Chofu, Tokyo 182-8585, Japan \\ 
$^{2}$Department of Applied Physics, Nagoya University,
 Nagoya 464-8603, Japan\\
}

\abst{
We microscopically study 
the effect of the magnetic field (Zeeman splitting) 
on the superconducting state 
in a model for quasi-one-dimensional 
organic superconductors (TMTSF)$_{2}X$. 
We investigate 
the competition between spin singlet and spin triplet pairings 
and the Fulde-Ferrell-Larkin-Ovchinnikov(FFLO) state 
by random phase approximation. 
While we studied the competition by comparison with the eigenvalue of 
the gap equation at a fixed temperature 
in our previous study (Phys. Rev. Lett. \textbf{102} (2009) 016403), 
here we obtain both 
the $T_c$ for each pairing state and a phase diagram in the 
$T$(temperature)-$h_z$(field)-$V_y$(strength of 
the charge fluctuation) space. 
The phase diagram shows that consecutive transitions from singlet 
pairing to the FFLO state and further to $S_z=1$ triplet pairing can 
occur 
upon increasing the magnetic field when $2k_{F}$ charge fluctuations 
coexist with $2k_{F}$ spin fluctuations. 
In the FFLO state, 
the singlet $d$-wave and $S_{z}=0$ triplet $f$-wave components are
strongly mixed especially when the charge fluctuations are strong.
}

\kword{ organic conductor, quasi-one-dimensional system, (TMTSF)$_{2}$X, 
 superconductivity, spin triplet pairing, FFLO state, Zeeman effect
}

\begin{document}
\maketitle


\section{Introduction\label{Introduction}}

The superconducting state 
of quasi-one-dimensional (Q1D) organic conductors 
(TMTSF)$_{2}X$ (TMTSF=tetramethyl-tetraselenafulvalene, 
$X$=PF$_{6}$, ClO$_{4}$ etc.) has been an issue of great interest. 
From the discovery of 
the first organic superconductor (TMTSF)$_{2}$PF$_{6}$, 
various studies have been performed both experimentally and 
theoretically. 
\cite{Jerome-Mazud-etal-TMTSF, Ishiguro-Yamaji-Saito, Lang-Muller, 
Coleman-Cohen-etal-TTF-TCNQ, Parkin-Engler-etal-BEDT-TTF, 
Bourbonnais-Jerome-review, Chem-Rev-104, JPSJ-75, Seo-Hotta-Fukuyama, 
Jerome-Chem-Rev, Lee-Brown-etal-JPSJ-Rev, Kuroki-JPSJ-Rev, 
Dupuis-Bourbonnais-etal-review} 
Previous studies for the NMR relaxation rate $1/T_{1}$ 
\cite{Takigawa-Yasuoka-etal-T1, Hasegawa-Fukuyama-T1} 
and the impurity effect 
\cite{Coulon-Delhaes-etal-impurity, Tomic-Jerome-etal-impurity, 
Choi-Chaikin-etal-impurity, Bouffard-Ribault-etal-impurity, 
Joo-Auban-etal-impurity1, Joo-Auban-etal-impurity2} 
have strongly suggested the possibility of anisotropic superconductivity  
where the nodes of the superconducting gap intersect the Fermi surface, 
although a thermal conductivity measurement has suggested 
the absence of nodes on the Fermi surface in (TMTSF)$_{2}$ClO$_{4}$. 
\cite{Belin-Behnia-thermal-conductivity}

Further experiments concerning the pairing symmetry 
have suggested the possibility that 
the pairing state in (TMTSF)$_{2}X$ may be even more fascinating. 
The NMR Knight shift measurements for (TMTSF)$_{2}$PF$_{6}$ 
and (TMTSF)$_{2}$ClO$_{4}$ have shown that the Knight shift is 
unchanged across the superconducting critical temperature $T_{c}$.  
\cite{Lee-Brown-etal-PF6-NMR-a, Lee-Chow-etal-PF6-NMR-b, 
Shinagawa-Wu-ClO4-NMR}
The upper critical field $H_{c2}$ 
for (TMTSF)$_{2}$PF$_{6}$ and (TMTSF)$_{2}$ClO$_{4}$ has been observed 
to exceed the Pauli paramagnetic limit $H_{{\rm P}}$. 
\cite{Lee-Naughton-etal-PF6-Hc2-PRL, 
Lee-Chaikin-etal-PF6-Hc2-PRB, Oh-Naughton-ClO4-Hc2}
These experiments suggest the possibility of spin triplet pairing and/or 
the Fulde-Ferrell-Larkin-Ovchinnikov (FFLO) state, 
\cite{Fulde-Ferrell, Larkin-Ovchinnikov}
in which the Cooper pairs formed 
as $( \textbf{\textit k}+\textbf{\textit Q}_{c}\uparrow, 
-\textbf{\textit k}+\textbf{\textit Q}_{c}\downarrow)$ 
have a finite center of mass momentum $\textbf{\textit Q}_{c}$.

Very recent experiments show more interesting results. 
The NMR experiment for (TMTSF)$_{2}$ClO$_{4}$ shows that 
the Knight shift changes across $T_{c}$ when the magnetic field is small,
but it is unchanged when a high magnetic field is applied. 
\cite{Shinagawa-Kurosaki-et-al} 
The $H_{c2}$ measurements for (TMTSF)$_{2}$ClO$_{4}$ 
show the possibility of two or three different pairing states. 
In the intermediate field regime, superconductivity is easily 
destroyed by tilting the magnetic field 
(between $H_{{\rm P}}$ and about 4T) 
from the conductive $a$-$b$ plane, while in the  
high field regime, superconductivity is sensitive to the broadening of 
the impurity scattering potential of the nonmagnetic impurity, 
namely, if the broadening of the impurity potential is large, 
the upturn curve of the critical temperature vanishes. 
\cite{Yonezawa-Kusaba-et-al-PRL, Yonezawa-Kusaba-et-al-JPSJ} 
These experiments suggest that spin singlet pairing 
occurs 
in the field regime lower than the Pauli limit, 
but spin triplet pairing and/or the FFLO state 
occurs 
in the higher field regime.

Theoretically, various studies on the pairing state in (TMTSF)$_{2}X$ 
have elucidated not only 
the unconventional pairing state with a nodal gap function 
\cite{Hasegawa-Fukuyama-T1,Kino-Kontani-TMTCF-FLEX,
Nomura-Yamada-TMTCF-TOP,Kuroki-Aoki-TMTCF-QMC, 
Kuroki-Tanaka-etal-TMTCF-QMC,Takigawa-Ichioka-et-al} 
but also 
the possibility of the spin triplet pairing 
\cite{Kuroki-Arita-Aoki,Tanaka-Kuroki,Kuroki-Tanaka,
Fuseya-Suzumura,Nickel-Duprat,
Lebed-FIDC,Lebed-Yamaji,Lebed-triplet,Lebed-Machida-Ozaki,
Shimahara-FIST,Vaccarella-Melo,Fuseya-Onishi-Kohno-et-al,
Belmechri-Abramovici-et-al-zeeman,Belmechri-Abramovici-et-al-orbital,
Aizawa-Kuroki-Tanaka,Aizawa-et-al-PRL}
and/or 
the FFLO state. 
\cite{Suzumura-Ishino,Machida-Nakanishi,Dupuis-Montambaux-Melo,
Dupuis-Montambaux,Miyazaki-Kishigi-Hasegawa,
Aizawa-et-al-PRL}  
In particular, we have previously shown 
that the spin triplet ``$f$-wave'' pairing 
can compete with the spin singlet ``$d$-wave'' pairing in Q1D systems 
\cite{comment-d-f-wave} 
when $2k_{F}$ spin fluctuations coexist with 
$2k_{F}$ charge fluctuations 
since the Fermi surface is disconnected in the $b$-direction. 
\cite{Kuroki-Arita-Aoki,Tanaka-Kuroki,Kuroki-Tanaka} 
In fact, the coexistence of $2k_{F}$ charge density wave and 
$2k_{F}$ spin density wave in the insulating phase 
has been observed by the diffuse X-ray 
scattering experiments in (TMTSF)$_{2}$PF$_{6}$. 
\cite{Pouget-Ravy,Kagoshima-Saso-et-al} 
A similar conclusion concerning the pairing state competition 
has been reached using the renormalization group technique. 
\cite{Fuseya-Suzumura,Nickel-Duprat} 
As a method for identifying spin-triplet $f$-wave pairing, 
tunneling spectroscopy   
\cite{Tanuma-Kuroki-PRB-66-094507, Tanuma-Tanaka-PRB-66-174502} 
via the mid gap Andreev resonant state 
\cite{Tanaka-Kashiwaya-PRL-74-3451, Kashiwaya-Tanaka-RPP-63-1641} 
and Josephson effect   
\cite{Asano-Tanaka-JPSJ-73-1922}
have been proposed.
In particular, the experiment of the proximity effect in the junctions 
with a diffusive normal metal is promising 
since an anomalous proximity effect with a zero-energy peak 
in the density of states, 
specific to spin-triplet superconductor junctions, has been predicted.  
\cite{
Tanaka-Kashiwaya-PRB-70-012507, Tanaka-Kashiwaya-PRB-71-094513, 
Tanaka-Asano-PRB-72-140503, Asano-Tanaka-PRL-96-097007, 
Tanaka-Nazarov-PRL-90-167003, Tanaka-Nazarov-PRB-69-144519}

There have also been various studies for the pairing state in the 
magnetic field.
The possibility of the FFLO state 
in finite magnetic field in a Q1D model for (TMTSF)$_{2}X$ 
has been suggested in several studies. 
\cite{Suzumura-Ishino,Machida-Nakanishi,Dupuis-Montambaux-Melo,
Dupuis-Montambaux,Miyazaki-Kishigi-Hasegawa}
The possibility of field-induced spin triplet pairing has also been 
discussed by a phenomenological theory and a renormalization technique. 
\cite{Lebed-FIDC,Lebed-Yamaji,Lebed-triplet,Lebed-Machida-Ozaki,
Shimahara-FIST,Vaccarella-Melo,Fuseya-Onishi-Kohno-et-al,
Belmechri-Abramovici-et-al-zeeman,Belmechri-Abramovici-et-al-orbital} 
Recently, we have microscopically studied the magnetic field effect 
on the pairing state in Q1D systems. 
We found that 
the $S_{z}=1$ triplet pairing mediated 
by $2k_{F}$ spin+$2k_{F}$ charge fluctuations is strongly enhanced 
by the magnetic field 
and showed the temperature-magnetic field phase diagram 
indicating the competition between 
the singlet and triplet pairings. \cite{Aizawa-Kuroki-Tanaka} 
We further found that the spin singlet, triplet, and FFLO states 
are closely competing, 
and the $S_{z}=0$ triplet component is strongly mixed with 
the singlet component in the FFLO state. 
There, the pairing state competition has been studied 
by comparing the eigenvalue of the linearized gap equation 
in the space of $V_y$ (strength of the charge fluctuation) 
and $h_z$ (magnetic field). \cite{Aizawa-et-al-PRL}

The FFLO state has recently been studied actively 
not only in Q1D but also in general systems. 
\cite{Casalbuoni-Nardulli,Matsuda-Shimahara} 
Previous theoretical studies have revealed various properties of 
the FFLO superconductivity from the viewpoint of 
(i) the orbital effect,  
\cite{Gruenberg-Gunther,Maki-Won,Shimahara-Rainer,
Tachiki-Takahashi-et-al,Houzet-Buzdin,Ikeda1,Ikeda2,Maniv-Zhuravlev,
Mizushima-Machida-Ichioka,Ichioka-Adachi-et-al,Klemm-Luther-Beasley,
Samokhin} 
(ii) the impurity effect, 
\cite{Takada,Agterberg-Yang,Adachi-Ikeda,Houzet-Mineev,Yanase-disorder}  
and 
(iii) the anisotropy of the system. 
\cite{Burkhardt-Rainer,
Shimahara-FFLO_direction-Q2D,Shimahara-FFLO_direction-kappa-ET,
Buzdin-Kachkachi,Vorontsov-Sauls-Graf,Suginishi-Shimahara-BETS,
Vorontsov-Graf,Shimahara-Moriwake,Kyker-Pickett} 
One of the interesting aspects of the FFLO state is  
parity mixing, i.e., even and odd parity pairings 
can be mixed to stabilize the FFLO state, 
which has been shown in phenomenological theories. 
\cite{Matsuo-Shimahara-Nagai,Shimahara2} 
Recent microscopic studies have also shown that 
the $S_{z}=0$ triplet pairing is mixed with singlet pairing 
in the FFLO state of the Hubbard model 
on the two-leg ladder-type lattice, 
\cite{Roux-White-Capponi-Poilblanc} 
the square lattice, 
\cite{Yanase-JPSJ-77-063705, Yokoyama-Onari-Tanaka-FFLO} 
and the Q1D extended Hubbard model. 
\cite{Aizawa-et-al-PRL} 
Yanase has pointed out that the parity mixing stabilizes the FFLO state, 
even in the vicinity of the quantum critical point,  
where the quasi-particle lifetime 
decreases owing 
to the scattering 
caused by spin fluctuations. \cite{Yanase-JPSJ-77-063705} 
In addition to 
these works, superconducting properties of the FFLO state 
have been studied theoretically. 
\cite{Vorontsov-Sauls-PRB-72-184501,Cui-Hu-PRB-73-214514,
Tanaka-Asano-PRL-98-077001}

Recent experiments strongly show the possibility of the FFLO state 
with an anisotropic gap function in CeCoIn$_{5}$.  
\cite{Radovan-Fortune-et-al,Bianchi-Movshovich-Capan-et-al,
Watanabe-Kasahara-et-al,Watanabe-Izawa-et-al,Cpan-Bianchi-et-al,
Correa-Murphy-et-al,Kakuyanagi-Sitoh-et-al,Kumagai-Saitoh-et-al,
Miclea-Nicklas-et-al,Gratens-Ferreira-et-al,Mitrovic'-Horvatic'-et-al,
Movshovich-Jaime-et-al,Hall-Palm-et-al,Bianchi-Movshovich-Oeschler-et-al,
Izawa-Yamaguchi-et-al,Aoki-Sakakibara-et-al,Vorontsov-Vekhter,
Martin-Agosta-et-al,Settai-Shishido-et-al,McCollam-Julian-et-al,
Young-Urbano-et-al}
Other candidate materials exhibiting the FFLO state are 
quasi-two-dimensional (Q2D) organic materials, 
such as $\lambda$-(BETS)$_{2}X$ 
(BETS=bisethylenedithio-tetraselenafulvalene, 
$X$=GaCl$_{4}$, 
\cite{Tanatar-Ishiguro-et-al} 
and FeCl$_{4}$ 
\cite{Uji-Shinagawa-et-al,Balicas-Brooks-et-al,Uji-Terashima-et-al}) 
and  
$\kappa$-(BEDT-TTF)$_{2}$Cu(NCS)$_{2}$ 
(BEDT-TTF=bisethylenedithio-tetrathiafulvalene), 
\cite{Manalo-Klein,Singleton-Symington-et-al,Lortz-Wang-et-al}
and also a Q1D one  (TMTSF)$_{2}$ClO$_{4}$ 
\cite{Shinagawa-Kurosaki-et-al,Yonezawa-Kusaba-et-al-PRL,
Yonezawa-Kusaba-et-al-JPSJ}. 
These materials have stimulated extensive studies in this field. 
The FFLO state attracts us not only in the field of superconductivity 
or superfluidity in condensed matter but also 
in the quantum chromodynamics 
\cite{Casalbuoni-Nardulli}
and the ultracold fermionic atom gas. 
\cite{Zwierlein-Schirotzek-et-al,Partridge-Li-et-al}

Given the above background, in this study, 
we investigate the pairing competition between 
the spin singlet and spin triplet pairings, and the FFLO state 
of the superconductivity mediated by 
spin and charge fluctuations 
in a Q1D extended Hubbard model for (TMTSF)$_{2}X$ 
by random phase approximation (RPA). 
\cite{Anderson-Brinkman, Nakajima-RPA, Miyake-Schmitt-Varma-RPA, 
 Scalapino-Loh-Hirsch-RPA, Shimahara-Takada-RPA} 
While the competition was studied 
only by comparison with the eigenvalue of the gap equation 
at a fixed temperature as indicated in ref. \citen{Aizawa-et-al-PRL}, 
here we calculate the superconducting temperature $T_{c}$  for each 
pairing state. 
This enables us to obtain a phase diagram in the $T$(temperature)-
$h_z$(field)-$V_y$(strength of the charge fluctuation) space, 
where we find that 
(i) consecutive transitions 
from singlet pairing to the FFLO state and further to 
$S_z=1$ triplet pairing can 
occur 
upon increasing the magnetic field in the vicinity of the SDW+CDW phase, 
and 
(ii) the enhancement of the charge fluctuations leads to 
a significant increase 
in parity mixing in the FFLO state, 
where the $S_{z}=0$ triplet/singlet component ratio 
in the gap function can be close to unity.

\section{Formulation\label{Formulation}}

\begin{figure}[!htb]
  \includegraphics[width=7.0cm]{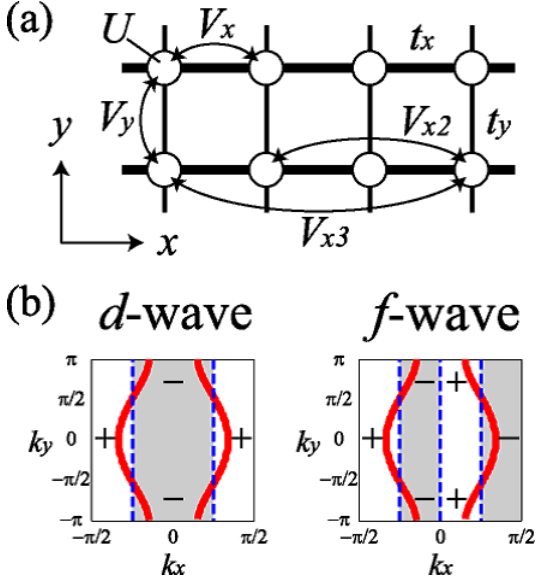}
  \caption{(Color online) 
  (a) Model adopted in this study. 
  (b) Schematic figure of the gap 
  for $d$-wave (left) and $f$-wave (right), 
  where blue dashed lines indicate the nodes of the gap, 
  and red solid curves indicate the disconnected Fermi surface.
  }
  \label{model}
 \end{figure}

The extended Hubbard model for (TMTSF)$_{2}X$ 
[Fig. \ref{model}(a)] 
that takes into account the Zeeman effect is given as 
\begin{eqnarray}
 H &=& 
  \sum_{i,j,\sigma} t_{ij\sigma} c_{i\sigma}^\dagger c_{j\sigma}
  +\sum_{i} U n_{i\uparrow}  n_{i\downarrow} 
\nonumber \\ & &
  +\sum_{i,j,\sigma,\sigma'} V_{ij} n_{i\sigma} n_{j\sigma'}. 
 \label{hamiltonian}
\end{eqnarray}
Here, $t_{ij\sigma}=t_{ij}+h_{z}{\rm sgn}(\sigma)\delta_{ij}$, where 
the hopping parameters $t_{ij}$ considered are the intrachain 
($a$-axis direction in (TMTSF)$_{2}X$) nearest-neighbor $t_x$ 
and the interchain ($b$-axis direction) nearest-neighbor $t_y$, 
$t_{x}=1.0$ being taken as the energy unit. 
$U$ is the on-site interaction, 
and $V_{ij}$ are the off-site interactions: 
$V_x$, $V_{x2}$, and $V_{x3}$ are the nearest-, next-nearest, and 
3rd-nearest-neighbor interactions within the chains, 
and $V_y$ is the interchain interaction. 
Note that we ignore the orbital effect, 
assuming that the magnetic field is applied parallel to 
the conductive $x$-$y$ plane,  
assuming a sufficiently large Maki parameter. 
(Since we neglect the orbital effect, 
the direction of the magnetic field within the $x$-$y$ plane 
is irrelevant within our approach.)  

The bare susceptibilities, 
consisting of bubble-type and ladder-type diagrams, are written as 
\begin{eqnarray}
 \chi_{0}^{\sigma \sigma}(k)
  &=&\frac{-1}{N}\sum_{q}
  \frac{f(\xi_{\sigma}(k+q))-f(\xi_{\sigma}(q))}
       {\xi_{\sigma}(k+q)-\xi_{\sigma}(q)},
  \label{a-chi0-para}
\\
 \chi_{0}^{+-}(k)
  &=&\frac{-1}{N}\sum_{q}
  \frac{f(\xi_{\sigma}(k+q))-f(\xi_{\bar{\sigma}}(q))}
       {\xi_{\sigma}(k+q)-\xi_{\bar{\sigma}}(q)},
  \label{a-chi0-pm}
\end{eqnarray}
where $\xi_{\sigma}(k)$ is the band dispersion 
that takes into account the Zeeman effect measured from 
the chemical potential $\mu$ 
and $f(\xi)$ is the Fermi distribution function.

Within RPA that takes into account the magnetic field parallel 
to the spin quantization axis $\hat{z}$, 
\cite{Aizawa-Kuroki-Tanaka, Aizawa-et-al-PRL} 
the longitudinal spin and charge susceptibilities are given by 
\begin{eqnarray}
 \chi_{\rm sp}^{zz}=\frac{1}{2}
   (\chi^{\uparrow \uparrow}+\chi^{\downarrow \downarrow}
  -\chi^{\uparrow \downarrow}-\chi^{\downarrow \uparrow}), 
\label{chispzz}
\\
 \chi_{\rm ch}=\frac{1}{2}
   (\chi^{\uparrow \uparrow}+\chi^{\downarrow \downarrow}
  +\chi^{\uparrow \downarrow}+\chi^{\downarrow \uparrow}), 
\label{chich}
\end{eqnarray}
where 
\begin{eqnarray}
 \chi^{\sigma \sigma}(k) &=&
  \left[ 1+\chi_{0}^{{\bar \sigma} {\bar \sigma}}(k) V(k)\right] 
  \chi_{0}^{\sigma \sigma}(k)/A(k), 
\label{a-chi-para}
\\
 \chi^{\sigma {\bar \sigma}}(k) &=&
  -\chi_{0}^{\sigma \sigma}(k)
  \left[U+V(k)\right]\chi_{0}^{{\bar \sigma} {\bar \sigma}}(k)/A(k), 
\label{a-chi-anti-para}
\\
 A(k) &=& \left[ 1+\chi_{0}^{\sigma \sigma}(k) V(k) \right]
    \left[ 1+\chi_{0}^{{\bar \sigma} {\bar \sigma}}(k) V(k)\right]
 \nonumber \\ & &
    -\left[U+V(k)\right]^{2}
    \chi_{0}^{\sigma \sigma}(k) 
    \chi_{0}^{{\bar \sigma} {\bar \sigma}}(k). 
\end{eqnarray}
The transverse spin susceptibility  
is given by
\begin{eqnarray}
 \chi_{\rm sp}^{+-}(k)=\frac{\chi_{0}^{+-}(k)}{1-U\chi_{0}^{+-}(k)},  
\label{chisppm}
\end{eqnarray}
where we ignore the off-site repulsions 
because it is difficult to treat the effect of the off-site repulsions 
on the ladder diagrams.

The pairing interactions from the bubble and ladder diagrams 
are given by
\begin{eqnarray}
 V^{\sigma \bar{\sigma}}_{\rm bub}(k) &=& 
 U+V(k)+\frac{U^{2}}{2}\chi_{\rm sp}^{zz}(k)
 \nonumber \\ & &
 -\frac{\left[ U+2V(k) \right]^{2}}{2}\chi_{\rm ch}(k), 
 \label{a-Vs-bub}
\\
 V^{\sigma \bar{\sigma}}_{\rm lad}(k) &=& U^{2}\chi_{\rm sp}^{+-}(k), 
 \label{a-Vs-lad}
\\
 V^{\sigma \sigma}_{\rm bub}(k) &=& 
 V(k)-2\left[ U+V(k) \right]V(k)\chi^{{\sigma} \bar{\sigma}}(k) 
 \nonumber \\ & & 
 -V(k)^{2}\chi^{{\sigma} {\sigma}}(k) 
 \nonumber \\ & & 
 -\left[ U+V(k) \right]^{2}\chi^{\bar{\sigma} \bar{\sigma}}(k), 
 \label{a-Vt-para-bub}\\
 V^{\sigma \sigma}_{\rm lad}(k) &=& 0. 
 \label{a-Vt-para-lad}
\end{eqnarray}
The linearized gap equation for Cooper pairs with 
the total momentum $2Q_{c}$ 
($Q_{c}$ represents the center of mass momentum) is given by 
\begin{eqnarray}
 \lambda^{\sigma \sigma'}_{Q_{c}} \varphi^{\sigma \sigma'}(k)
  = \frac{1}{N}\sum_{q}
  [V^{\sigma \sigma'}_{\rm bub}(k-q)+V^{\sigma \sigma'}_{\rm lad}(k+q)]
  \nonumber \\  
 \times
  \frac{ f(\xi_{\sigma}(q_{+})) 
        -f(-\xi_{\sigma'}(-q_{-}))}
       {\xi_{\sigma}(q_{+})+\xi_{\sigma'}(-q_{-})}
  \varphi^{\sigma \sigma'}(q), 
 \label{gap-eq}
\end{eqnarray}
where $q_{\pm}=q \pm Q_{c}$, 
$\varphi^{\sigma \sigma'}(k)$ is the gap function, 
and $\lambda^{\sigma \sigma'}_{Q_{c}}$ is the eigenvalue of this 
linearized gap equation.
The center of mass momentum $\textbf{\textit{Q}}_{c}$, 
which gives the maximum value of 
$\lambda^{\sigma \bar{\sigma}}_{Q_{c}}$, lies in the $x$-direction, 
\cite{Shimahara-FFLO_direction-Q2D, Shimahara-FFLO_direction-kappa-ET, 
Yokoyama-Onari-Tanaka-FFLO} 
while $\lambda^{\sigma \sigma}_{Q_{c}}$ takes its maximum  
at $\textbf{\textit{Q}}_{c}=(0,0)$ 
because the electrons with the same spin can be paired as 
$(k\sigma, -k\sigma)$ for all $k$.

We define the singlet and $S_{z}=0$ triplet components 
of the gap function in the opposite spin pairing channel as 
\begin{eqnarray}
 \varphi_{{\rm SS}}(k)&=&
  \frac{ \varphi^{\uparrow \downarrow}(k)
  -\varphi^{\downarrow \uparrow}(k) }{2}, 
\nonumber \\
 \varphi_{{\rm ST}^{0}}(k)&=&
  \frac{ \varphi^{\uparrow \downarrow}(k)
   +\varphi^{\downarrow \uparrow}(k) }{2}.
 \label{eq-phis-phit}
\end{eqnarray} 
In our calculation, 
the spin singlet and triplet components of the gap function 
in the FFLO state are essentially the $d$-wave and $f$-wave, 
respectively, 
as schematically shown in Fig. \ref{model}(b); 
thus, we write the singlet ($S_{z}=0$ triplet) component of the FFLO gap 
$\varphi_{{\rm SS}}$($\varphi_{{\rm ST}^{0}}$) 
in eq. (\ref{eq-phis-phit}) as 
$\varphi_{{\rm SS}d}$ ($\varphi_{{\rm ST}f^{0}}$), 
where SS$d$(ST$f^{0}$) stands for spin singlet $d$-wave 
(spin triplet $f$-wave with $S_{z}=0$) pairing. 
The eigenvalue of each pairing state is determined as follows. 
$\lambda^{\sigma \bar{\sigma}}_{Q_{c}}$ with 
$\textbf{\textit{Q}}_{c}=(0,0)$ gives 
the eigenvalue of the singlet $d$-wave pairing $\lambda_{{\rm SS}d}$ 
($S_z=0$ triplet $f$-wave pairing $\lambda_{{\rm ST}f^{0}}$) 
$\varphi_{{\rm ST}f^{0}}=0$ ($\varphi_{{\rm SS}d}=0$), 
while $\lambda^{\sigma \bar{\sigma}}_{Q_{c}}$ 
with $\textbf{\textit{Q}}_{c} \ne (0,0)$ gives $\lambda_{{\rm FFLO}}$. 
$\lambda^{\sigma \sigma}_{Q_{c}}$ with 
$\textbf{\textit{Q}}_{c}=(0,0)$ gives the eigenvalue for 
the spin triplet $f$-wave pairing with $S_{z}=+1$ ($S_{z}=-1$) 
$\lambda_{{\rm ST}f^{+1}}$ ($\lambda_{{\rm ST}f^{-1}}$).
The above-mentioned results of the determination of the eigenvalues 
are listed in Table \ref{pairing-sort-table}. 

\begin{table}[htbp]
\begin{center}
\caption{ Results of the determination of the eigenvalue of the linearized 
gap equation $\lambda^{\sigma \sigma'}_{\textbf{\textit{Q}}_{c}}$. }
\label{pairing-sort-table}
\begin{tabular}{|p{0.7cm}|p{3.0cm}|p{3.0cm}|}
\hline 
 & Center of mass momentum and paired spins 
 & Pairing symmetry or SC state \\ \hline\hline
      & $\textbf{\textit{Q}}_{c}=0$, $\sigma \ne \sigma'$ 
      & singlet $d$-wave ($\lambda_{{\rm SS}d}$)
        for $\varphi_{{\rm ST}f^{0}}\left( k \right)=0$
\\ \cline{2-3}
      & $\textbf{\textit{Q}}_{c}=0$, $\sigma \ne \sigma'$ 
      & $S_{z}=0$ triplet $f$-wave ($\lambda_{{\rm ST}f^{0}}$)
        for $\varphi_{{\rm SS}d}\left( k \right)=0$
\\ \cline{2-3}
$\lambda^{\sigma \sigma'}_{\textbf{\textit{Q}}_{c}}$ 
      & $\textbf{\textit{Q}}_{c}=0$, $\sigma = \sigma'$ 
      & $S_{z}=\pm 1$ triplet $f$-wave ($\lambda_{{\rm ST}f^{\pm 1}}$)
\\ \cline{2-3}
      & $\textbf{\textit{Q}}_{c} \ne 0$, $\sigma \ne \sigma'$ 
      & FFLO state ($\lambda_{\rm FFLO}$)
 \\ \cline{2-3}
      & $\textbf{\textit{Q}}_{c} \ne 0$, $\sigma = \sigma'$ 
      & not dominant state \\ \hline
\end{tabular}
\end{center}
\end{table}

Although RPA is quantitatively insufficient for discussing 
the absolute value of $T_{c}$, 
we expect this approach to be valid for studying 
the competition between different pairing symmetries.
In this paper, we fix the hopping parameters as 
$t_{x}=1.0$ and $t_{y}=0.2$, and  
the electron-electron interactions as 
$U=1.7$, $V_{x}=0.9$, $V_{x2}=0.45$, and $V_{x3}=0.1$, 
and vary $V_{y}$. 
Since the dimerization of TMTSF molecules is very small 
in (TMTSF)$_{2}X$ compounds, 
we ignore the dimerization and 
fix the band filling as $n=1.5$ (3/4 filling), 
where $n=$ number of electrons/number of sites. 
1024$\times$128 $k$-point meshes are taken, where we 
take a large number of $k_x$ meshes  
since the center of mass momentum $\textbf{\textit{Q}}_{c}$, 
which gives the maximum value of the FFLO state, lies in 
the $x$-direction.

\section{Results\label{Results}}

\subsection{ Center of mass momentum and the gap function}

\begin{figure}[!htb]
 \includegraphics[width=8.0cm]{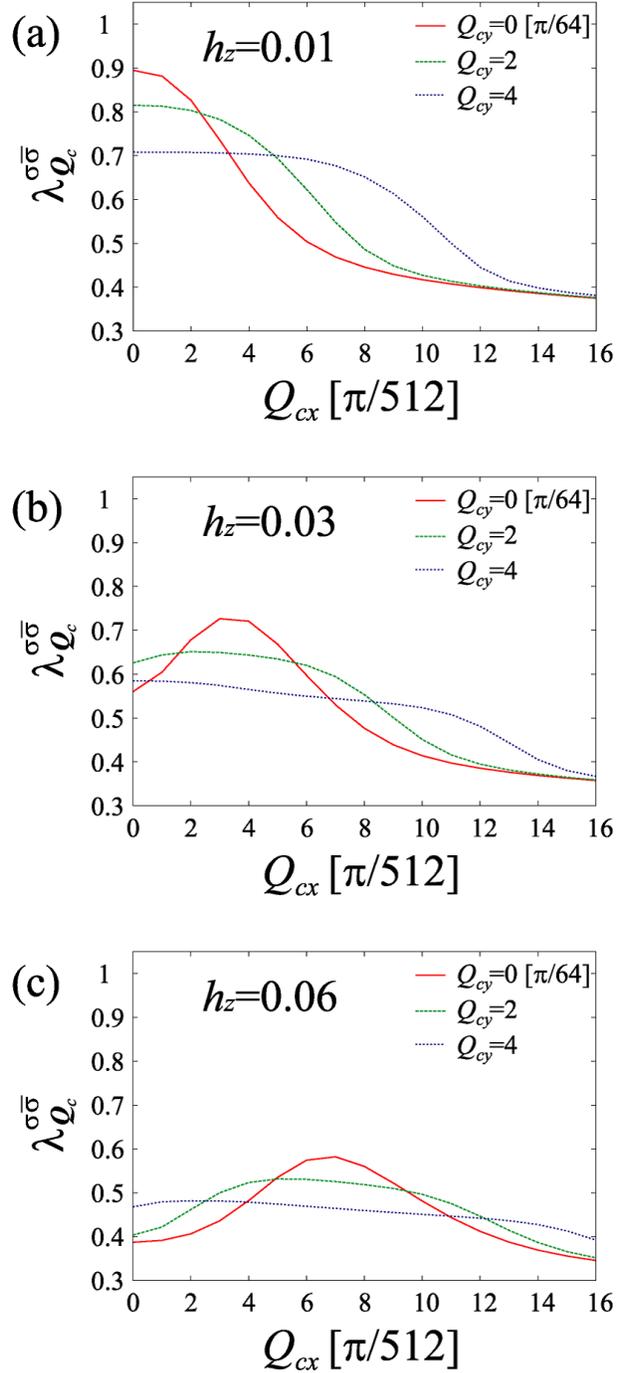}
 \caption{(Color online) 
 ${\textit{Q}}_{cx}$-dependence of the eigenvalue in the 
 opposite-spin pairing channel, 
 $\lambda_{\textbf{\textit{Q}}_{c}}^{\sigma \bar{\sigma}}$, 
 for (a) $h_{z}=0.01$, (b) $h_{z}=0.03$, and (c) $h_{z}=0.06$ 
 at $V_{y}=0.35$. }
 \label{fig2}
\end{figure}

In this section, we study the nature of the FFLO state in our model.
Let us first study the center of mass momentum 
at which the FFLO state is most stabilized.
The optimum $\textbf{\textit{Q}}_{c}$ that most stabilizes the FFLO state 
can be determined as $\textbf{\textit{Q}}_{c}$ at which the 
eigenvalue of the gap equation is maximized.
In the following results, we set the interchain off-site interaction 
as $V_{y}=0.35$. 
$\lambda_{\textbf{\textit{Q}}_{c}}^{\sigma \bar{\sigma}}$ with 
$\textbf{\textit{Q}}_{c}
=\left( \textit{Q}_{cx}, \textit{Q}_{cy} \right)$ 
are given in units of $\pi/512$ for the $x$-direction 
and $\pi/64$ for the $y$-direction. 
Figure \ref{fig2} shows the eigenvalue of the linearized gap equation 
in the opposite-spin pairing channel 
$\lambda_{\textbf{\textit{Q}}_{c}}^{\sigma \bar{\sigma}}$ 
as a function of the $x$-component of the center of mass momentum 
$\textit{Q}_{cx}$ for various $\textit{Q}_{cy}$. 
When the magnetic field is small ($h_{z}=0.01$), 
the pairing state with $\textbf{\textit{Q}}_{c}=\left( 0, 0 \right)$ 
dominates over other finite momentum states, 
as seen in Fig. \ref{fig2}(a). 
For a larger magnetic field ($h_{z}=0.03$),  
a finite momentum pairing state with 
$\textit{Q}_{cx}=3$ and $\textit{Q}_{cy}=0$ dominates 
over other states in the opposite-spin pairing channel, 
as shown in Fig. \ref{fig2}(b). 
When the magnetic field is increased up to $h_{z}=0.06$, 
a finite momentum pairing state with 
$\textit{Q}_{cx}=7$ and $\textit{Q}_{cy}=0$ 
is the most dominant, but the eigenvalue 
$\lambda_{\textbf{\textit{Q}}_{c}}^{\sigma \bar{\sigma}}$ itself 
decreases, 
as shown in Fig. \ref{fig2}(c). 
Further studying other $h_z$ cases, 
we find that the most dominant center of mass momentum lies in the 
$x$-direction, 
\cite{Shimahara-FFLO_direction-Q2D, Shimahara-FFLO_direction-kappa-ET, 
Yokoyama-Onari-Tanaka-FFLO} 
and the magnitude of the center of mass momentum increases 
with increasing magnetic field. 
\cite{Yokoyama-Onari-Tanaka-FFLO}

\begin{figure}[!htb]
 \includegraphics[width=7.0cm]{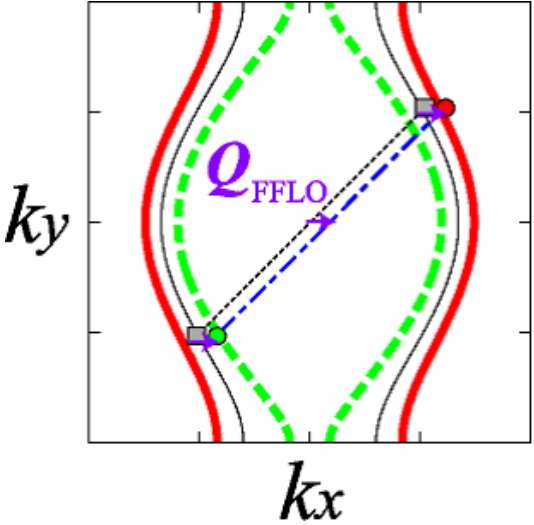}
 \caption{(Color online) 
 The small purple arrow in the $k_{x}$-direction 
 denoting $\textbf{\textit{Q}}_{\rm FFLO}$ schematically shows 
 the direction of the center of mass momentum vector 
 $\textbf{\textit{Q}}_{c}$ in the FFLO state. 
 The black thin solid cuarves schemetically represent 
 the Fermi surface in zero field.  
 The red thick solid (green thick dashed) curves schematically 
 represent the Fermi surface split by the Zeeman effect. 
 The red and green filled circles are the particle 
 on each Fermi surface in the presence of the field, 
 and the gray filled squares are the particle in zero field. }
 \label{fig3}
\end{figure}

The direction of the center of mass momentum vector 
$\textbf{\textit{Q}}_{c}$ can be understood from 
the Fermi surface split by the Zeeman effect shown in Fig. \ref{fig3}. 
The electron pair part, i.e., the particle-particle susceptibility, 
in the linearized gap equation is rewritten as 
\begin{eqnarray}
& &  \frac{ f(\xi_{\sigma}(q+Q_{c})) 
        -f(-\xi_{\sigma'}(-q+Q_{c}))}
       {\xi_{\sigma}(q+Q_{c})+\xi_{\sigma'}(-q+Q_{c})} 
\nonumber \\ 
&=& \frac{1}{\beta}\sum_{\varepsilon_{n}} 
G_{\sigma }\left(  q+Q_{c}, i \varepsilon_{n} \right)
G_{\sigma'}\left( -q+Q_{c},-i \varepsilon_{n} \right), 
\,\,\,\,\,\,\,\,\,\,
 \label{GG}
\end{eqnarray}
where $\xi_{\sigma}(-k+Q_{c})$ is the same as $\xi_{\sigma}(k-Q_{c})$
since $\xi_{\sigma}(k)=\xi_{\sigma}(-k)$ is satisfied. 
If the $\sigma$ spin electron energy at the wave vector $q+Q_c$ and 
the $\sigma'$ spin electron energy at the wave vector $-q+Q_c$ 
are close to the Fermi energy, 
the denominator is small and 
eq. (\ref{GG}) can take a large value. 
For a quasi-one-dimensional system, the number of 
wave vectors $q$ that satisfies such a condition becomes the largest  
when the vector $Q_c$ is in the $k_x$ direction

\begin{figure}[!htb]
 \includegraphics[width=8.5cm]{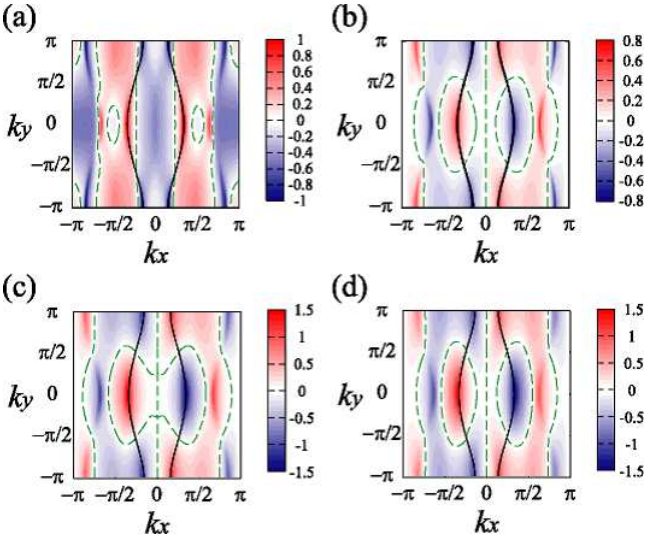}
 \caption{(Color online) 
 Gap function for 
 (a) singlet component, 
 (b) $S_{z} = 0$ triplet component in the FFLO state with 
 $Q_{cx}=3$ and $Q_{cy}=0$, 
 (c) $S_{z} = +1$ triplet state, and  
 (d) $S_{z} = -1$ triplet state, 
 where black solid curves represent the Fermi surface, and 
 the green dashed curves represent the nodes of the gap. 
 The parameters are $h_{z}=0.03$, $V_{y}=0.35$, and $T=0.012$. }
 \label{fig4}
\end{figure}

Next, we study the gap functions normalized 
by the maximum value of the singlet component gap function 
in the FFLO state. 
We set the parameters as $h_{z}=0.03$, $V_{y}=0.35$, and $T=0.012$, 
where the FFLO state 
which has the finite center of mass momentum as $(Q_{cx}, Q_{cy})=(3, 0)$ 
is the most dominant, as described later. 
Note that the $S_{z} = \pm 1$ triplet pairings always have 
the maximum value of the eigenvalue 
$\lambda_{\textbf{\textit{Q}}_{c}}^{\sigma \sigma}$ 
at $\textbf{\textit{Q}}_{c}=\left( 0, 0 \right)$, 
as mentioned previously. 
As shown in Figs. \ref{fig4}(a) and \ref{fig4}(b), 
the singlet component of the gap function in the FFLO state 
is the $d$-wave and  
the $S_{z}=0$ triplet component is the $f$-wave.
The maximum value of the $S_{z}=0$ triplet gap component 
in the FFLO state almost reaches unity. 
Thus, the singlet $d$-wave component and 
the $S_{z}=0$ triplet $f$-wave component strongly mix 
in this FFLO state. 
The gap function in the $S_{z}=\pm 1$ triplet pairings has 
the $f$-wave form shown in Figs. \ref{fig4}(c) and \ref{fig4}(d). 

The appearance of the $d$-wave gap in the singlet component 
and the $f$-wave gap in the $S_{z}=0$ triplet component in the 
FFLO state is understood as follows.
In zero field, 
the singlet $d$-wave pairing mediated by the $2k_{F}$ spin fluctuations 
is favored in the Q1D Hubbard model,
namely,  
the large  pairing interaction due to 
the $2k_{F}$ spin fluctuations stabilizes 
the spin singlet $d$-wave pairing. 
\cite{Kuroki-Arita-Aoki,Kino-Kontani-TMTCF-FLEX, 
Nomura-Yamada-TMTCF-TOP,Kuroki-Aoki-TMTCF-QMC,
Kuroki-Tanaka-etal-TMTCF-QMC}
Moreover, 
the coexistence of 
$2k_{F}$ charge fluctuations, which is induced by 
the second-nearest-neighboring repulsive interaction,
favors the triplet $f$-wave pairing 
in the Q1D extended Hubbard model at quarter filling. 
\cite{Kuroki-Arita-Aoki,Tanaka-Kuroki,Kuroki-Tanaka,
Fuseya-Suzumura,Nickel-Duprat}
The reason why the spin triplet $f$-wave pairing can compete with 
the spin singlet $d$-wave pairing in the Q1D extended Hubbard model is 
(i) the contribution of the $2k_{F}$ charge fluctuations 
in the pairing interaction  enhances 
the spin triplet $f$-wave pairing and suppresses 
the spin singlet $d$-wave pairing, and 
(ii) $f$ and $d$-wave pairings have the same number of gap nodes 
intersecting the Fermi surface due to the disconnectivity 
of the Fermi surface (quasi-one-dimensionality).
The above mechanism is valid 
even in the presence of the magnetic field, but more importantly, 
the spin triplet $f$-wave pairing mediated by 
the $2k_{F}$ spin + $2k_{F}$ charge fluctuations 
can be enhanced by applying the magnetic field 
since the bubble-type diagram enhanced by the field 
contributes to the pairing interaction without being 
paired with the bubble-type diagram, which is suppressed by the field. 
\cite{Aizawa-Kuroki-Tanaka}
Actually, our previous work shows 
a clear correlation between the $S_{z}=0$ triplet ratio 
in the FFLO state 
and the ratio of the eigenvalue between the $S_{z}=0$ triplet and 
singlet pairings obtained by the formulation of separating 
the singlet and $S_{z}=0$ triplet channels. 
\cite{Aizawa-et-al-PRL} 
From the above, 
we can understand not only 
the appearance of the $d$-wave ($f$-wave) gap 
in the singlet ($S_{z}=0$ triplet) component of the opposite-spin 
pairing channel and the $f$-wave gap in the parallel-spin 
pairing channel,  
but also the 
large parity mixing of the singlet and $S_{z}=0$ triplet components 
in the FFLO state.

\begin{figure}[!htb]
 \includegraphics[width=8.0cm]{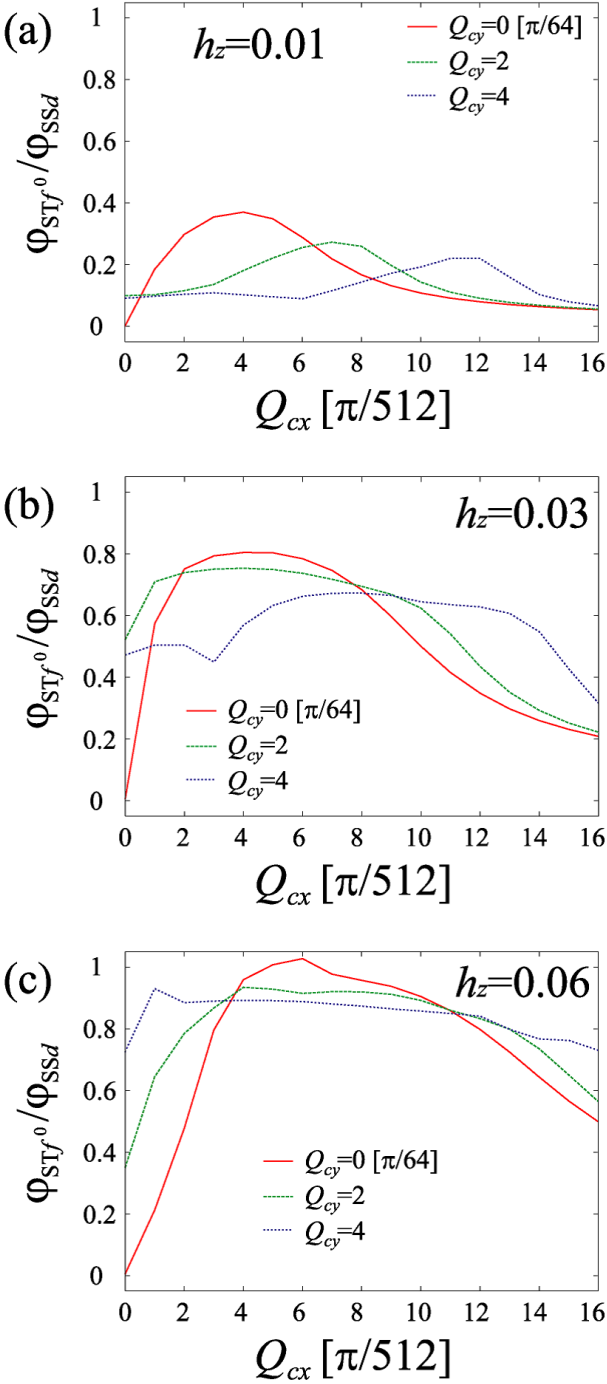}
 \caption{(Color online) 
 ${\textit{Q}}_{cx}$-dependence of the parity mixing 
 in the opposite-spin pairing channel, 
 $\varphi_{{\rm ST}f^{0}}/\varphi_{{\rm SS}d}$, 
 for (a) $h_{z}=0.01$, (b) $h_{z}=0.03$, and (c) $h_{z}=0.06$ 
 at $V_{y}=0.35$. }
 \label{fig5}
\end{figure}

Figure \ref{fig5} shows 
the parity mixing $\varphi_{{\rm ST}f^{0}}/\varphi_{{\rm SS}d}$ 
in the opposite-spin pairing channel as a function of 
the $x$-component of the center of mass momentum $Q_{cx}$ 
for several $Q_{cy}$.  
Note that we need to bear the $\textit{ \textbf{Q}}_{c}$ dependence 
of the eigenvalue 
$\lambda_{\textbf{\textit{Q}}_{c}}^{\sigma \bar{\sigma}}$, 
as shown in Fig. \ref{fig2}, in order to see 
the $\textit{ \textbf{Q}}_{c}$ dependence of the parity mixing rate 
because the most dominant pairing state in the opposite-spin pairing  
state is determined by the value of 
$\lambda_{\textbf{\textit{Q}}_{c}}^{\sigma \bar{\sigma}}$. 
For instance, 
we have seen in Fig. \ref{fig2}(a) that the singlet $d$-wave pairing, 
i.e., the opposite-spin 
pairing state with $( Q_{cx}, Q_{cy} )=( 0, 0 )$, 
is the most dominant in the small magnetic field regime. 
In Fig. \ref{fig5}(a), 
the parity mixing rate for $h_{z}=0.01$  
$\varphi_{{\rm ST}f^{0}}/\varphi_{{\rm SS}d}$ is zero  
at $( Q_{cx}, Q_{cy} )=( 0, 0 )$.  
Therefore, no $S_{z}=0$ triplet $f$-wave component is present 
in this pairing state, and 
the opposite-spin pairing channel is a purely spin singlet $d$-wave.
For $h_{z}=0.03$, we have seen in Fig. \ref{fig2}(b) that 
the FFLO state with $Q_{cx}=3$ and $Q_{cy}=0$ is dominant. 
As shown in Fig. \ref{fig5}(b), the parity mixing rate for 
$Q_{cx}=3$ and $Q_{cy}=0$ takes a large value 
$\varphi_{{\rm ST}f^{0}}/\varphi_{{\rm SS}d} \simeq 0.8$. 
For $h_{z}=0.06$, where 
the FFLO state with $Q_{cx}=7$ and $Q_{cy}=0$ is dominant 
(Fig. \ref{fig2}(c)), 
the parity mixing rate increases, $i.e.$, 
$\varphi_{{\rm ST}f^{0}}/\varphi_{{\rm SS}d} \simeq 1.0$, 
as shown in Fig. \ref{fig5}(c), 
which means that the singlet $d$-wave component and 
the $S_{z}=0$ triplet $f$-wave component are strongly  mixed 
in this FFLO state (provided this state is actually realized).
The strong parity mixing in the FFLO state can be understood 
as a consequence of the breaking of the spacial inversion symmetry in the 
superconducting state. 
Previous theoretical studies have shown that 
the parity mixing with the singlet and triplet pairings 
stabilizes the FFLO state more when 
only the singlet component is considered.

\subsection{Temperature dependence}

\begin{figure}[!htb]
 \includegraphics[width=8.0cm]{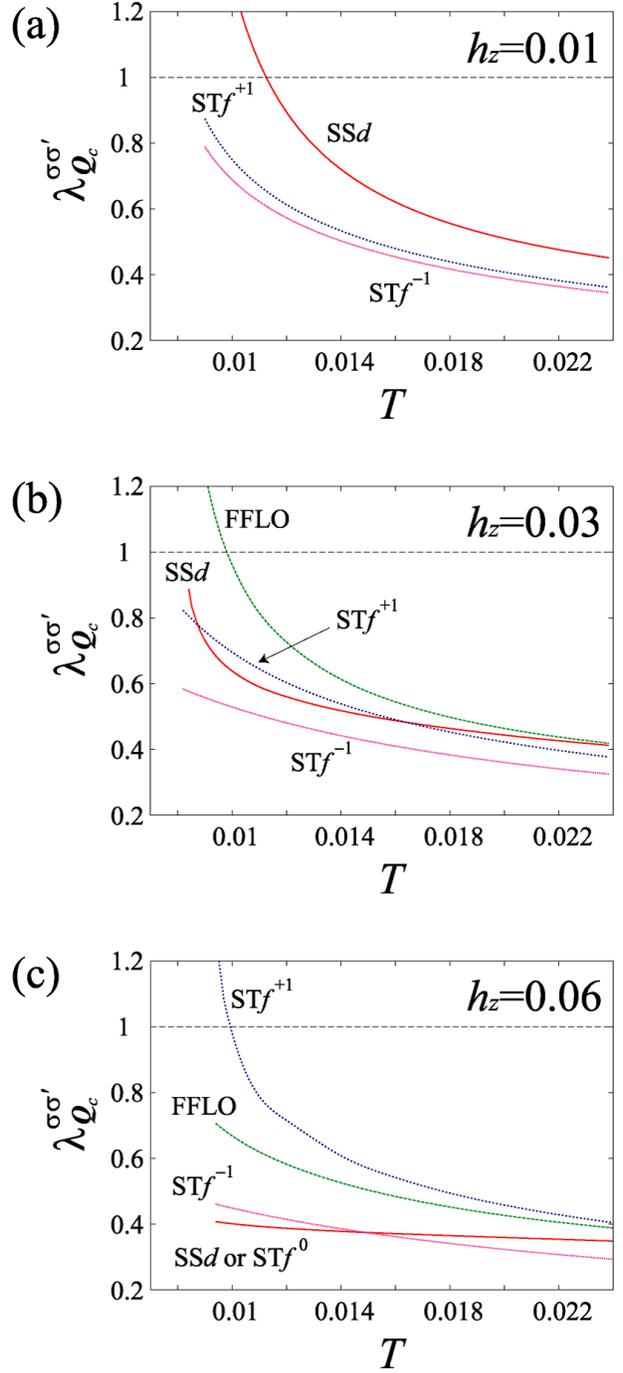}
 \caption{(Color online) 
 Eigenvalue of the linearized gap equation, 
 $\lambda_{\textbf{\textit{Q}}_{c}}^{\sigma \sigma'}$, 
 plotted as a function of the temperature $T$ 
 for (a) $h_{z}=0.01$, (b) $h_{z}=0.03$, and (c) $h_{z}=0.06$ 
 with $V_{y}=0.35$. 
 Note that SS$d$ and ST$f^{\pm 1}$ have 
 $\textbf{\textit{Q}}_{c}=\textbf{0}$, and 
 the FFLO state has a finite $\textbf{\textit{Q}}_{c}$ that 
 maximizes the eigenvalue of the opposite-spin channel 
 in Fig. \ref{fig2}.}
 \label{fig6}
\end{figure}

Next, we investigate the temperature dependence of the 
eigenvalue, $\lambda_{\textbf{\textit{Q}}_{c}}^{\sigma \sigma'}$, 
in both the opposite- and parallel-spin pairing states. 
We have confirmed that 
the center of mass momentum that most stabilizes 
the FFLO state is unchanged upon lowering the temperature for a fixed 
magnetic field.
For  $h_{z}=0.01$, 
the eigenvalue 
$\lambda_{{\rm SS}d} = 
\lambda_{\textbf{\textit{Q}}_{c}={\textbf 0}}^{\sigma \bar{\sigma}}$
of the spin singlet $d$-wave pairing reaches unity, 
as shown in Fig. \ref{fig6}(a). 
In this small magnetic field regime, 
the FFLO state is absent, as shown in Fig. \ref{fig2}(a).
For $h_{z}=0.03$, 
the singlet $d$-wave pairing is suppressed and 
the eigenvalue 
$\lambda_{{\rm FFLO}} = 
\lambda_{\textbf{\textit{Q}}_{c} \ne {\textbf 0}}^{\sigma \bar{\sigma}}$
of the FFLO state with $Q_{cx}=3$ and $Q_{cy}=0$ 
reaches unity as seen in Fig. \ref{fig6}(b). 
For $h_{z}=0.06$, 
the FFLO state with $Q_{cx}=7$ and $Q_{cy}=0$ does not 
develop much upon lowering the temperature, 
while the eigenvalue 
$\lambda_{{\rm ST}f^{+1}} = 
\lambda_{\textbf{\textit{Q}}_{c}={\textbf 0}}^{\sigma \sigma}$
for the $S_{z}=1$ triplet $f$-wave state 
reaches unity, as shown in Fig. \ref{fig6}(c). 
The eigenvalue of the 
singlet $d$-wave and $S_{z}=-1$ triplet $f$-wave pairings 
remains small even in the low temperature regime. 

\subsection{Calculated phase diagram}

\begin{figure}[!htb]
 \includegraphics[width=8.0cm]{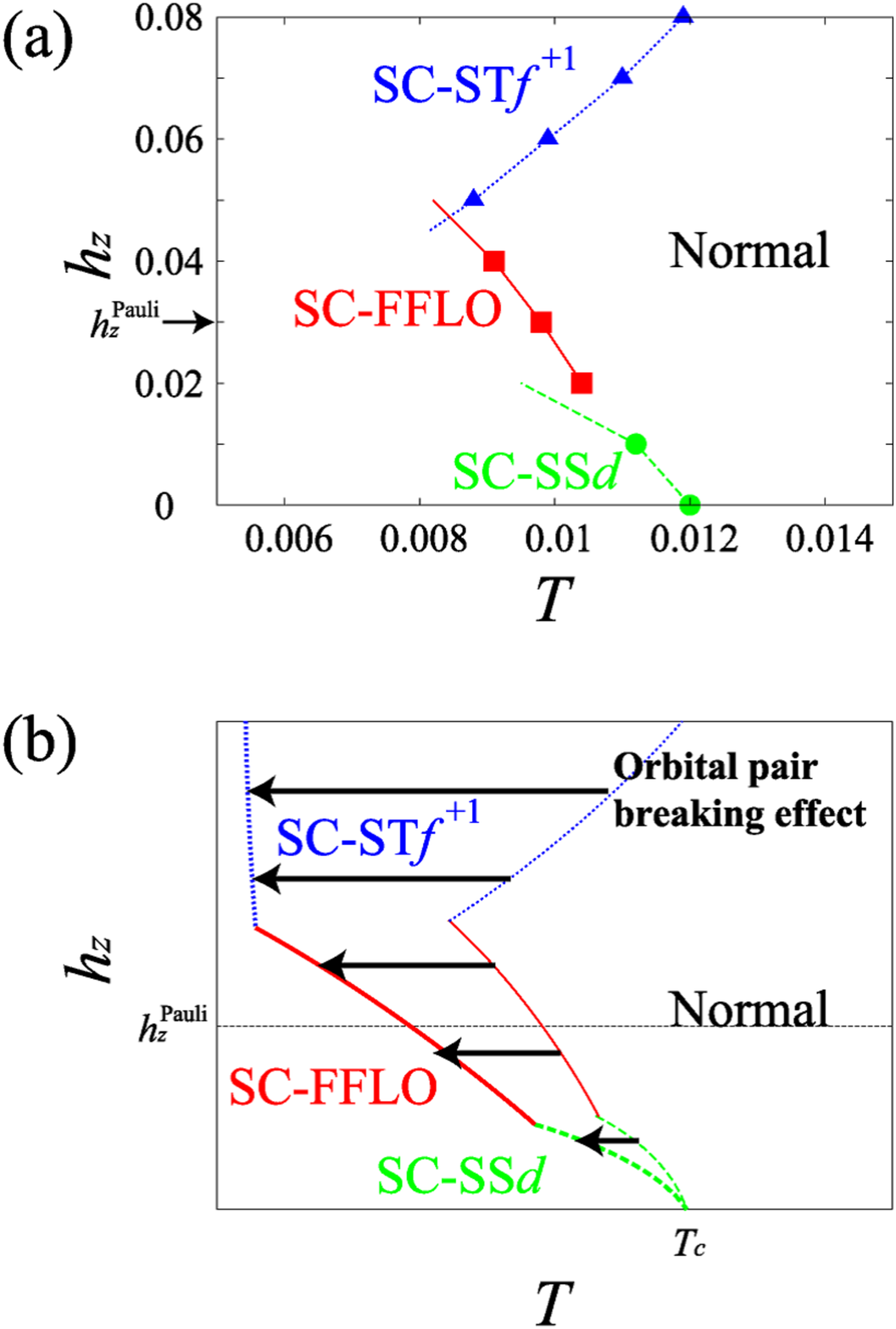}
 \caption{(Color online) 
 (a) Calculated phase diagram in $h_{z}$-$T$ space for $V_{y}=0.35$, 
 where 
 the green dashed line indicates the $T_{c}$ for 
 the spin singlet $d$-wave, 
 the red solid line represents that for the FFLO state, and 
 the blue dotted line indicates that 
 for the $S_{z}=1$ spin triplet $f$-wave. 
 The spin singlet $d$-wave is omitted as ${\rm SS}d$ and 
 the $S_{z}=1$ spin triplet $f$-wave is ${\rm ST}f^{+1}$. 
 The same notation is used in Figs. \ref{fig8} and \ref{fig9}. 
 (b) Schematic figure of the orbital pair breaking effect on 
 the superconducting phase diagram in $T$-$h_{z}$ space, 
 where black solid arrows schematically represent 
 the orbital pair breaking effect. }
 \label{fig7}
\end{figure}

We now obtain a phase diagram in the temperature $T$ versus 
the magnetic field $h_{z}$ space 
for several values of interchain off-site interaction $V_{y}$
(which controls the strength of the charge fluctuations). 
Figure \ref{fig7}(a) shows a plot of the critical temperature $T_{c}$ 
against the magnetic field $h_{z}$ for $V_{y}=0.35$, 
where the $2k_{F}$ charge fluctuations are slightly weaker 
than the $2k_{F}$ spin fluctuations. 
The critical temperature in zero field is 
$T_{c} \simeq 0.012$ and the   
estimated value of Pauli's paramagnetic field is 
$h_{z}^{\rm P} \simeq 0.03$. 
We see that a consecutive transition from singlet pairing 
to the FFLO state and further to $S_{z}=1$ triplet pairing 
occurs 
upon increasing the magnetic field. 

This consecutive pairing transition can be understood as follows. 
It is known that the FFLO state can be 
stabilized by the quasi-one-dimensionality, 
namely, the nesting of the Fermi surface.
\cite{Shimahara-FFLO_direction-Q2D, Shimahara-FFLO_direction-kappa-ET, 
Yokoyama-Onari-Tanaka-FFLO} 
Thus, the quasi-one-dimensionality of the present model is 
one of the origins of the transition from the $d$-wave 
to the FFLO state. 
The origin of the pairing transition from the FFLO state to 
the $S_{z}=1$ triplet pairing is understood by 
our previous study, where we have shown that 
the triplet pairing 
due to the coexisting $2k_{F}$ spin and $2k_{F}$ charge fluctuations 
is strongly enhanced by the 
direct contribution of the unpaired bubble diagram 
enhanced by the field. 
\cite{Aizawa-Kuroki-Tanaka} 

Here, we emphasize that we ignore the orbital pair breaking 
effect in this study 
because our aim in this work is to study the competition between 
the singlet, FFLO, and triplet pairings in the case when 
the magnetic field is applied in the conductive plane, 
i.e., the $a$-$b$ plane of (TMTSF)$_{2}X$. 
For discussing the above pairing competition, 
the Zeeman splitting effect is essential for the FFLO state; thus, 
we ignore the orbital pair breaking effect at the beginning.  
Although this effect is small in applying the magnetic field 
parallel to the conductive plane, 
the orbital pair breaking effect is present in actual materials.  
Furthermore, previous studies have shown that 
the orbital pair breaking effect is important 
in  discussing the FFLO superconductivity. 
\cite{Gruenberg-Gunther,Maki-Won,Shimahara-Rainer,
Tachiki-Takahashi-et-al,Houzet-Buzdin,Ikeda1,Ikeda2,Maniv-Zhuravlev,
Mizushima-Machida-Ichioka,Ichioka-Adachi-et-al,Klemm-Luther-Beasley,
Samokhin} 

As shown in Fig. \ref{fig7}(a),   
there seems to be a reentrance from the superconducting state to 
another superconducting state intervened by the normal state. 
However, it is much more reasonable to 
consider that this reentrance does not actually 
occur owing 
the presence of the  orbital pair breaking effect. 
If this effect is taken into account, 
not only the FFLO state but also 
the singlet and triplet pairing states 
should strongly be suppressed upon increasing the magnetic field. 
Figure \ref{fig7}(b) shows a schematic figure of the 
effect of the orbital pair breaking, where the 
$T_c$ obtained (without the orbital effect) 
in Fig. \ref{fig7}(a) (thin curve)
is suppressed down to the thick curve.
The thick curve in Fig. \ref{fig7}(b) 
is reminiscent of the experimental $T$-$H$ phase diagram  
\cite{Lee-Naughton-etal-PF6-Hc2-PRL,Lee-Chaikin-etal-PF6-Hc2-PRB, 
Oh-Naughton-ClO4-Hc2,
Yonezawa-Kusaba-et-al-PRL,Yonezawa-Kusaba-et-al-JPSJ} 
in that the $T_{c}$ curve makes an upturn from  nearly above 
the Pauli limit.

\begin{figure}[!htb]
 \includegraphics[width=8.0cm]{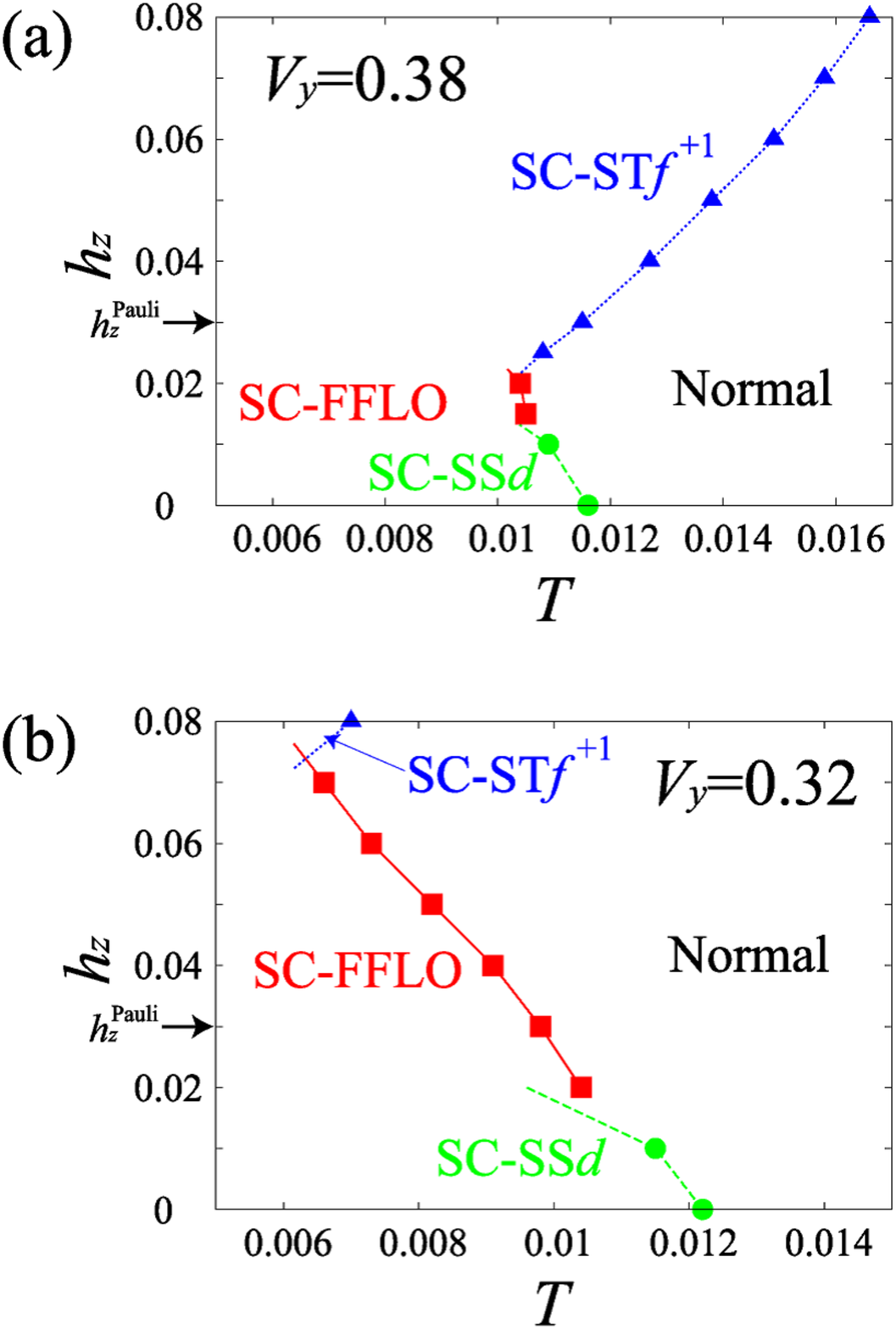}
 \caption{(Color online) 
 Calculated phase diagram in $h_{z}$-$T$ space for 
 (a) $V_{y}=0.38$ and (b) $V_{y}=0.32$, 
 where the notation is the same as that in Fig. \ref{fig7}(a). }
 \label{fig8}
\end{figure}

Next, we study the effect of the interchain interaction $V_{y}$ 
on the phase diagram in the temperature $T$ versus 
the magnetic field $h_{z}$ space. 
Figure \ref{fig8}(a) shows the critical temperature $T_{c}$ 
at each magnetic field $h_{z}$ for $V_{y}=0.38$. 
The magnetic field at which the transition from the FFLO state 
to the $S_{z}=1$ triplet $f$-wave pairing occurs is smaller 
than that in the $V_{y}=0.35$ case. 
The critical temperature $T_{c}$ for $V_{y}=0.32$ is shown 
in Fig. \ref{fig8}(b), 
which shows that 
the magnetic field at which the FFLO state gives way to the  
$S_{z}=1$ triplet $f$-wave pairing is 
larger than those in the previous phase diagrams shown 
in Figs. \ref{fig7}(a) and \ref{fig8}(a). 
The difference between the two phase diagrams is due to the 
fact that the $2k_F$ charge fluctuations enhance the triplet $f$-wave 
pairing; thus, the FFLO state appears only in a small parameter regime 
in between the $d$- and $f$-wave pairings.

\begin{figure}[!htb]
 \includegraphics[width=8.0cm]{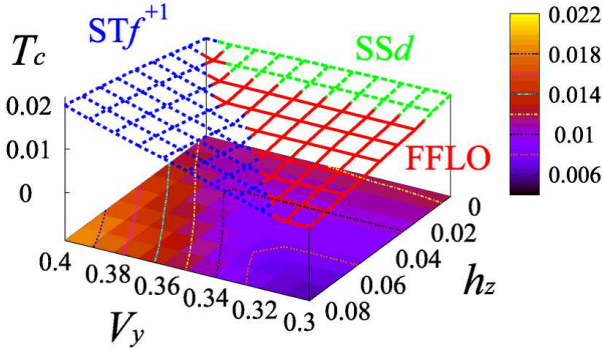}
 \caption{(Color online) 
 The critical temperature is shown in the $V_{y}$-$h_{z}$ plane. 
 The value of $T_{c}$ is plotted in the vertical axis 
 and represented by contours. 
 Green dashed lines represent the critical temperature 
 for the singlet $d$-wave pairing, 
 red solid lines are the $T_{c}$ for the FFLO state, 
 and the blue dotted lines are the $T_{c}$ for 
 the $S_{z}=1$ spin triplet $f$-wave pairing, respectively. 
 }
 \label{fig9}
\end{figure}

Summarizing the above-mentioned features, 
we show the phase diagram in  $T$-$V_y$-$h_z$ space in Fig. \ref{fig9}. 
When $V_{y}$ is small and thus the $2k_{F}$ spin fluctuations 
are dominant over the $2k_{F}$ charge fluctuations, 
$T_{c}$ decreases and 
the transition from the spin singlet $d$-wave to the FFLO state 
occurs 
upon increasing $h_{z}$. 
In this FFLO state, the strong parity mixing with 
the spin singlet $d$-wave component and 
the $S_{z}=0$ spin triplet $f$-wave component 
occurs. 
In the large $V_{y}$ regime, the $2k_{F}$ charge fluctuations 
compete with the $2k_{F}$ spin fluctuations, 
and the consecutive pairing state transition 
from the spin singlet $d$-wave to 
the FFLO state and further to the $S_{z}=1$ spin triplet $f$-wave 
upon increasing the $h_{z}$ at the critical temperature $T_{c}$ 
occurs. 
The $T_{c}$ enhancement of the $S_{z}=1$ spin triplet $f$-wave pairing 
in the large $h_{z}$ regime can be understood by 
our previous work. \cite{Aizawa-Kuroki-Tanaka}

\section{Conclusion\label{Conclusion}}

We have studied the competition between spin singlet, triplet, 
and FFLO superconductivities in a model for (TMTSF)$_{2}X$ 
by applying the RPA method and solving the linearized gap equation 
within the weak coupling theory. 
We find the following: 

(i) consecutive pairing transitions 
from singlet pairing to the FFLO state and further to 
$S_{z}=1$ triplet pairing can 
occur 
upon increasing the magnetic field 
in the vicinity of the SDW+CDW coexisting phase. 
 
(ii) in the FFLO state,
the $S_{z}=0$ spin triplet pairing component is mixed with 
the spin singlet pairing component, 
thus resulting in a large parity mixing.

Recent experiments for (TMTSF)$_{2}$ClO$_4$ suggest differences in 
superconducting properties 
in the low and high field regimes. 
The Knight shift study shows 
the presence of low field and high field pairing states, 
where the former is the spin singlet pairing and 
the latter is the FFLO state or the spin triplet pairing. 
\cite{Shinagawa-Kurosaki-et-al}
The upper critical field studies have shown that 
only the clean sample, or more strictly, samples where the broadening of 
the nonmagnetic impurity is small, 
exhibits an upturn of the critical temperature curve 
in the high field parallel to the $a$ axis regime above 4T; 
thus, the high field superconducting state is sensitive 
to the impurity content or the anisotropy of the impurity 
scattering potential. 
\cite{Yonezawa-Kusaba-et-al-JPSJ}
Between 4T and the Pauli limit around 2.5T, 
there seems to be a different high field pairing state, 
in which superconductivity is stable against the impurities, 
but it is very sensitive to the tilt of the magnetic field out of 
the $a$-$b$ plane. 
The bottom line of these experiments is that 
there may be three kinds of pairing states, 
i.e., one low field state and two high field states. 
The correspondence between these experimental observations and 
the present study is not clear at the present stage, 
but the appearance of the three kinds of pairing states 
is indeed intriguing. 
It would be interesting to further investigate experimentally 
the possibility and nature of two kinds of high field pairing states.

One point that should be mentioned for 
(TMTSF)$_{2}$ClO$_{4}$ in particular is the presence of 
the anion ordering 
with the modulation wave vector 
$\textbf{\textit Q}_{\rm AO} = \left( 0, \pi/b \right)$, 
which takes place near $T_{\rm AO} \simeq$ 24K 
when slowly cooled.
Recent studies show that the anion ordering potential ($V_{\rm AO}$) 
is around $0.02 t_{x}$. 
\cite{Yoshino-Shodai-SM-133-55, Lebed-Ha-PRB-71-132504}
The anion ordering leads to a folding of the Brillouin zone 
in the $k_{y}$-direction ($b$-direction), and in that case, 
the $d$-wave (and also $f$-wave in the same sense) 
gap can become nodeless 
because the folded Fermi surface becomes disconnected 
near the nodes of the gap, as has been suggested by Shimahara. 
\cite{Shimahara-PRB-61-R14938} 
This effect is neglected in our present study, 
and its effect on the pairing symmetry competition is 
an interesting future problem.

Another point to be mentioned is that in the present study, 
we do not take account of the 
retardation effect. By taking account of this effect, $i.e.$, 
the frequency dependence of the gap function, 
we can discuss the odd-frequency pairing state. 
\cite{Berezinskii-JETP-Lett-20-287,Balatsky-Abrahams-PRB-45-13125,
Bergeret-Volkov-RMP-77-1321}
It has been shown  that odd-frequency pairing can be realized 
in a certain quasi-one-dimensional lattice. 
\cite{Shigeta-Onari}
In particular, in the presence of non-uniformity, 
the odd-frequency pairing amplitude is ubiquitously generated. 
\cite{Tanaka-Golubov-PRL-98-037003,Tanaka-Golubov-PRL-99-037005,
Tanaka-Tanuma-PRB-76-054533,Tanaka-Asano-PRB-77-220504R,
Yokoyama-Tanaka-PRB-78-012508,Tanuma-Hayashi-PRL-102-11703}
It is a future interesting problem to study the possible existence of 
odd-frequency pairing in quasi-one-dimensional organic superconductors.

\section*{Acknowledgment}
We acknowledge S. Yonezawa for valuable discussions. 
This work is supported by Grants-in-Aid for Scientific Research from the
Ministry of Education, Culture, Sports, Science and Technology of
Japan, and from the Japan Society for the Promotion of Science.
Part of the calculation has been performed at the 
facilities of the Supercomputer Center, 
ISSP, University of Tokyo.



\begin{thebibliography}{999} 

\bibitem{Jerome-Mazud-etal-TMTSF} 
 D. J${\rm \acute{e}}$rome, A. Mazud,  M. Ribault, and K. Bechgaard: 
 J. Phys. Lett. (Paris) \textbf{41} (1980) L95. 

\bibitem{Ishiguro-Yamaji-Saito} 
 T. Ishiguro, K. Yamaji, and G. Saito:  
 \textit{Organic Superconductors} 
 (Springer-Verlag,  Heidelberg,  1998) 2nd ed.
 
\bibitem{Lang-Muller}
 M. Lang and J. M${\rm \ddot{u}}$ller:  
 arXiv:cond-mat/0302157 
 published in 
 \textit{The Physics of Superconductors - Vol.2} 
 (Springer-Verlag,  Heidelberg,  2003). 

\bibitem{Coleman-Cohen-etal-TTF-TCNQ}
 L. B. Coleman, M. J. Cohen, D. J. Sandman, F. G. Yamagishi, 
 A. F. Gatiro, and A. J. Heeger: 
 Solid State Commun. \textbf{12} (1973) 1125.

\bibitem{Parkin-Engler-etal-BEDT-TTF}
 S. S. P. Parkin, E. M. Engler, R. R. Schumaker, R. Lagier, 
 V. Y. Lee, J. C. Scott, and R. L. Greene: 
 Phys. Rev. Lett. \textbf{50} (1983) 270. 

\bibitem{Bourbonnais-Jerome-review}
 C. Bourbonnais and D. J${\rm \acute{e}}$rome: 
 arXiv:cond-mat/9903101 
 published in 
 \textit{Advances in Synthetic Metals, Twenty years of Progress 
 in Science and Technology} 
 (Elsevier, New York, 1999). 

\bibitem{Chem-Rev-104}
 For a recent review, Chem. Rev. (2004) \textbf{104}, 
 special issue on Molecular Conductors. 

\bibitem{JPSJ-75}
 For a recent review, J. Phys. Soc. Jpn. (2006) \textbf{75}, 
 special topics on Organic Conductors. 

\bibitem{Seo-Hotta-Fukuyama} 
 H. Seo,  C. Hotta, and H. Fukuyama: 
 Chem. Rev. \textbf{104} (2004) 5005. 

\bibitem{Jerome-Chem-Rev} 
 D. J${\rm \acute{e}}$rome: 
 Chem. Rev. \textbf{104} (2004) 5565. 

\bibitem{Lee-Brown-etal-JPSJ-Rev}
 I. J. Lee, S. E. Brown, and M. J. Naughton: 
 J. Phys. Soc. Jpn. \textbf{75} (2006) 051011. 

\bibitem{Kuroki-JPSJ-Rev}
 K. Kuroki: 
 J. Phys. Soc. Jpn. \textbf{75} (2006) 051013. 

\bibitem{Dupuis-Bourbonnais-etal-review}
 N. Dupuis, C. Bourbonnais, and J. C. Nickel: 
 Low. Temp. Phys. \textbf{32} (2006) 380. 


\bibitem{Takigawa-Yasuoka-etal-T1}
 M. Takigawa, H. Yasuoka, and G. Saito: 
 J. Phys. Soc. Jpn. \textbf{56} (1987) 873.

\bibitem{Hasegawa-Fukuyama-T1}
 Y. Hasegawa and H. Fukuyama: 
 J. Phys. Soc. Jpn. \textbf{56} (1987) 877.

\bibitem{Bouffard-Ribault-etal-impurity}
 S. Bouffard, M. Ribault, R. Brusetti, D. J${\rm \acute{e}}$rome, 
 and K. Bechgaard: 
 J. Phys. C \textbf{15} (1982) 2951. 

\bibitem{Coulon-Delhaes-etal-impurity}
 C. Coulon, P. Delha${\rm \acute{e}}$s, J. Amiell, J. P. Manceau, 
 J. M. Fabre, and L. Giral: 
 J. Phys. (Paris) \textbf{43} (1982) 1721.  

\bibitem{Choi-Chaikin-etal-impurity}
 M. Y. Cohi, P. M. Chaikin, S. Z. Huang, P. Haen, E. M. Engler, 
 and R. L. Greene: 
 Phys. Rev. B \textbf{25} (1982) 6208. 
 
\bibitem{Tomic-Jerome-etal-impurity}
 S. Tomic, D. J${\rm \acute{e}}$rome, D. Mailly, M. Ribault, 
 and K. Bechgaard: 
 J. Phys. (Paris) \textbf{44} (1983) C3-1075.  

\bibitem{Joo-Auban-etal-impurity1}
 N. Joo, P. Auban-Senzier, C. R. Pasquier, P. Monod, 
 D. J${\rm \acute{e}}$rome, and K. Bechgaard: 
 Eur. Phys. J. B \textbf{40} (2004) 43. 

\bibitem{Joo-Auban-etal-impurity2}
 N. Joo, P. Auban-Senzier, C. R. Pasquier, 
 D. J${\rm \acute{e}}$rome, and K. Bechgaard: 
 Europhys. Lett. \textbf{72} (2005) 645. 

\bibitem{Belin-Behnia-thermal-conductivity}
 S. Belin and K. Behnia: 
 Phys. Rev. Lett. \textbf{79} (1997) 2125. 

\bibitem{Lee-Brown-etal-PF6-NMR-a}
 I. J. Lee, S. E. Brown, W. G. Clark, M. J. Strouse, M. J. Naughton, 
 W. Kang, and P. M. Chaikin: 
 Phys. Rev. Lett. \textbf{88} (2002) 017004. 

\bibitem{Lee-Chow-etal-PF6-NMR-b}
 I. J. Lee, D. S. Chow, W. G. Clark, M. J. Strouse, M. J. Naughton, 
 P. M. Chaikin, and S. E. Brown: 
 Phys. Rev. B \textbf{68} (2003) 092510. 

\bibitem{Shinagawa-Wu-ClO4-NMR}
 J. Shinagawa, W. Wu, P. M. Chaikin, W. Kang, W. Yu, F. Zhang, 
 Y. Kurosaki, C. Parker, and S. E. Brown: 
 J. Low Temp. Phys. \textbf{142} (2007) 227. 

\bibitem{Lee-Naughton-etal-PF6-Hc2-PRL}
 I. J. Lee, M. J. Naughton, G. M. Danner, and P. M. Chaikin: 
 Phys. Rev. Lett. \textbf{78} (1997) 3555. 

\bibitem{Lee-Chaikin-etal-PF6-Hc2-PRB}
 I. J. Lee, P. M. Chaikin, and M. J. Naughton: 
 Phys. Rev. B \textbf{62} (2000) R14669. 

\bibitem{Oh-Naughton-ClO4-Hc2}
 J. I. Oh and M. J. Naughton: 
 Phys. Rev. Lett. \textbf{92} (2004) 067001.  


\bibitem{Fulde-Ferrell} 
 P. Fulde and R. A. Ferrell: 
 Phys. Rev. \textbf{135} (1964) A550. 

\bibitem{Larkin-Ovchinnikov} 
 A. I. Larkin and Yu. N. Ovchinnikov: 
 Zh. Eksp. Teor. Fiz. \textbf{47} (1964) 1136
 [Sov. Phys. JETP \textbf{20} (1965) 762]. 


\bibitem{Shinagawa-Kurosaki-et-al}
 J. Shinagawa, Y. Kurosaki, F. Zhang, C. Parker, S. E. Brown, 
 D. J${\rm \acute{e}}$rome, J. B. Christensen, and K. Bechgaard: 
 Phys. Rev. Lett. \textbf{98} (2007) 147002.  

\bibitem{Yonezawa-Kusaba-et-al-PRL}
 S. Yonezawa, S. Kusaba, Y. Maeno, P. Auban-Senzier, C. Pasquier, 
 K. Bechgaard, and D. J${\rm \acute{e}}$rome: 
 Phys. Rev. Lett. \textbf{100} (2008) 117002. 

\bibitem{Yonezawa-Kusaba-et-al-JPSJ}
 S. Yonezawa, S. Kusaba, Y. Maeno, P. Auban-Senzier, C. Pasquier, 
 and D. J${\rm \acute{e}}$rome: 
 J. Phys. Soc. Jpn. \textbf{77} (2008) 054712. 



\bibitem{Kino-Kontani-TMTCF-FLEX}
 H. Kino and H. Kontani: 
 J. Low. Temp. Phys. \textbf{177} (1999) 317. 

\bibitem{Kuroki-Aoki-TMTCF-QMC} 
 K. Kuroki and H. Aoki: 
 Phys. Rev. B \textbf{60} (1999) 3060.  

\bibitem{Nomura-Yamada-TMTCF-TOP}
 T. Nomura and K. Yamada: 
 J. Phys. Soc. Jpn. \textbf{70} (2001) 2694. 

\bibitem{Kuroki-Tanaka-etal-TMTCF-QMC} 
 K. Kuroki, Y. Tanaka, T. Kimura, and R. Arita: 
 Phys. Rev. B \textbf{69} (2004) 214511.  

\bibitem{Takigawa-Ichioka-et-al} 
 M. Takigawa, M. Ichioka, K. Kuroki, Y. Asano, and Y. Tanaka: 
 Phys. Rev. Lett. {\bf 97} (2006) 187002.


\bibitem{Kuroki-Arita-Aoki}
 K. Kuroki, R. Arita, and H. Aoki: 
 Phys. Rev. B \textbf{63} (2001) 094509. 

\bibitem{Tanaka-Kuroki}
 Y. Tanaka and K. Kuroki: 
 Phys. Rev. B \textbf{70} (2004) 060502. 

\bibitem{Kuroki-Tanaka}
 K. Kuroki and Y. Tanaka: 
 J. Phys. Soc. Jpn. \textbf{74} (2005) 1694. 

\bibitem{Fuseya-Suzumura} 
 Y. Fuseya and Y. Suzumura: 
 J. Phys. Soc. Jpn. {\bf 74} (2005) 1263.

\bibitem{Nickel-Duprat} 
 J. C. Nickel, R. Duprat, C. Bourbonnais, and N. Dupuis: 
 Phys. Rev. Lett. {\bf 95} (2005) 247001.


\bibitem{Lebed-FIDC}
 A. G. Lebed: 
 JETP Lett. \textbf{44} (1986) 114. 

\bibitem{Lebed-Yamaji} 
 A. G. Lebed and K. Yamaji: 
 Phys. Rev. Lett. \textbf{80} (1998) 2697.   

\bibitem{Lebed-triplet} 
 A. G. Lebed: 
 Phys. Rev. B \textbf{59} (1999) R721.   

\bibitem{Lebed-Machida-Ozaki} 
 A. G. Lebed, K. Machida, and M. Ozaki: 
 Phys. Rev. B \textbf{62} (2000) R795.   

\bibitem{Shimahara-FIST}
 H. Shimahara: 
 J. Phys. Soc. Jpn. \textbf{69} (2000) 1966. 

\bibitem{Vaccarella-Melo} 
 C. D. Vaccarella and  C. A. R. S${\rm \acute{a}}$ de Melo: 
 Physica C \textbf{341-348} (2000) 293. 

\bibitem{Fuseya-Onishi-Kohno-et-al}
 Y. Fuseya, Y. Onishi, H. Kohno, and K. Miyake: 
 J. Phys.: Condens. Matter \textbf{14} (2002) L655. 

\bibitem{Belmechri-Abramovici-et-al-zeeman} 
 N. Belmechri, G. Abramovici, M. H${\rm \acute{e}}$ritier, S. Haddad, 
 and S. Charfi-Kaddour: 
 Europhys. Lett. \textbf{80} (2007) 37004.   

\bibitem{Belmechri-Abramovici-et-al-orbital} 
 N. Belmechri, G. Abramovici, and M. H${\rm \acute{e}}$ritier: 
 Europhys. Lett. \textbf{82} (2008) 47009. 


\bibitem{Aizawa-Kuroki-Tanaka} 
 H. Aizawa, K. Kuroki, and Y. Tanaka: 
 Phys. Rev. B \textbf{77} (2008) 144513. 


\bibitem{Aizawa-et-al-PRL} 
 H. Aizawa, K. Kuroki, T. Yokuyama and Y. Tanaka: 
 Phys. Rev. Lett. \textbf{102} (2009) 016403. 


\bibitem{Suzumura-Ishino} 
 Y. Suzumura and K. Ishino: 
 Prog. Theor. Phys. \textbf{70} (1983) 654.  

\bibitem{Machida-Nakanishi} 
 K. Machida and H. Nakanishi: 
 Phys. Rev. B \textbf{30} (1984) 122. 

\bibitem{Dupuis-Montambaux-Melo} 
 N. Dupuis, G. Montambaux, and C. A. R. S${\rm \acute{a}}$ de Melo: 
 Phys. Rev. Lett. \textbf{70} (1993) 2613.   

\bibitem{Dupuis-Montambaux} 
 N. Dupuis and G. Montambaux: 
 Phys. Rev. B \textbf{49} (1994) 8993.   

\bibitem{Miyazaki-Kishigi-Hasegawa} 
 M. Miyazaki, K. Kishigi, and Y. Hasegawa: 
 J. Phys. Soc. Jpn. \textbf{68} (1999) 3794. 


\bibitem{comment-d-f-wave} 
Strictly speaking, 
the gap function in the left (right) figure of Fig. \ref{model}(b) 
is not the $d$-wave ($f$-wave) in the sense that 
it does not have an angular momentum equal to two (three). 
In this paper, however, 
we call these states as the ``$d$-wave'' and ``$f$-wave'' 
in a broad sense (as in the previous studies) 
in that the gap changes the sign 
as $+ - + -$ (``$d$'')  or $+ - + - + -$ $(``f'')$ along 
the Fermi surface.


 \bibitem{Pouget-Ravy}
 J. P. Pouget and S. Ravy: 
 J. Phys. I (Paris) \textbf{6} (1996) 1501. 

 \bibitem{Kagoshima-Saso-et-al}
 S. Kagoshima, Y. Saso, M. Maesato, R. Kondo, and T. Hasegawa: 
 Solid State Commun. \textbf{110} (1999) 479. 


 \bibitem{Tanuma-Kuroki-PRB-66-094507} 
 Y. Tanuma, K. Kuroki, Y. Tanaka, R. Arita, S. Kashiwaya, and H. Aoki: 
 Phys. Rev. B \textbf{66} (2002) 094507. 

 \bibitem{Tanuma-Tanaka-PRB-66-174502} 
 Y. Tanuma, Y. Tanaka, K. Kuroki, and S. Kashiwaya:  
 Phys. Rev. B \textbf{66} (2002) 174502. 

 \bibitem{Tanaka-Kashiwaya-PRL-74-3451}
 Y. Tanaka and S. Kashiwaya: 
 Phys. Rev. Lett. \textbf{74} (1995) 3451. 

 \bibitem{Kashiwaya-Tanaka-RPP-63-1641} 
 S. Kashiwaya and Y. Tanaka: 
 Rep. Prog. Phys. \textbf{63} (2000) 1641. 

 \bibitem{Asano-Tanaka-JPSJ-73-1922}
 Y. Asano, Y. Tanaka, Y. Tanuma, K. Kuroki, and H. Tsuchiura:  
 J. Phys. Soc. Jpn. \textbf{73} (2004) 1922.  

\bibitem{Tanaka-Nazarov-PRL-90-167003}
 Y. Tanaka, Y. V. Nazarov, and S. Kashiwaya:  
 Phys. Rev. Lett. \textbf{90} (2003) 167003.  

\bibitem{Tanaka-Nazarov-PRB-69-144519}
 Y. Tanaka, Y. V. Nazarov, A. A. Golubov, and S. Kashiwaya:  
 Phys. Rev. B \textbf{69} (2004) 144519. 

\bibitem{Tanaka-Kashiwaya-PRB-70-012507} 
 Y. Tanaka and S. Kashiwaya:  
 Phys. Rev. B \textbf{70} (2004) 012507. 

\bibitem{Tanaka-Kashiwaya-PRB-71-094513} 
 Y. Tanaka, S. Kashiwaya, and T. Yokoyama:  
 Phys. Rev. B \textbf{71} (2005) 094513. 

\bibitem{Tanaka-Asano-PRB-72-140503} 
 Y. Tanaka, Y. Asano, A. A. Golubov, and S. Kashiwaya:   
 Phys. Rev. B \textbf{72} (2005) 140503(R). 

\bibitem{Asano-Tanaka-PRL-96-097007} 
 Y. Asano, Y. Tanaka, and S. Kashiwaya:  
 Phys. Rev. Lett. \textbf{96} (2006) 097007. 


\bibitem{Casalbuoni-Nardulli}
 For a review, see 
 R. Casalbuoni and G. Nardulli: 
 Rev. Mod. Phys. \textbf{76} (2004) 263. 

\bibitem{Matsuda-Shimahara}
 For a review, see 
 Y. Matsuda and H. Shimahara: 
 J. Phys. Soc. Jpn. \textbf{76} (2007) 051005. 



\bibitem{Gruenberg-Gunther} 
 L. W. Gruenberg and L. Gunther: 
 Phys. Rev. Lett. \textbf{16} (1966) 996. 

\bibitem{Maki-Won}
 K. Maki and H. Won: 
 Czech. J. Phys. \textbf{46} (1996) Suppl. S2, 1035. 

\bibitem{Shimahara-Rainer} 
 H. Shimahara and D. Rainer: 
 J. Phys. Soc. Jpn. \textbf{66} (1997) 3591. 

\bibitem{Tachiki-Takahashi-et-al} 
 M. Tachiki, S. Takahashi, P. Gegenwart, M. Weiden, M. Lang, C. Geibel, 
 F. Steglich, R. Modler, C. Paulsen, and Y. ${\rm \bar{O}}$nuki: 
 Z. Phys. B \textbf{100} (1996) 369. 

\bibitem{Houzet-Buzdin} 
 M. Houzet and A. Buzdin: 
 Phys. Rev. B \textbf{63} (2001) 184521. 

\bibitem{Ikeda1} 
 R. Ikeda: 
 Phys. Rev. B \textbf{76} (2007) 134504.  

\bibitem{Ikeda2} 
 R. Ikeda: 
 Phys. Rev. B \textbf{76} (2007) 054517. 

\bibitem{Maniv-Zhuravlev} 
 T. Maniv and V. Zhuravlev: 
 Phys. Rev. B \textbf{77} (2008) 134511.  

\bibitem{Mizushima-Machida-Ichioka} 
 T. Mizushima, K. Machida, and M. Ichioka: 
 Phys. Rev. Lett. \textbf{95} (2005) 117003.  

\bibitem{Ichioka-Adachi-et-al} 
 M. Ichioka, H. Adachi, T. Mizushima, and K. Machida: 
 Phys. Rev. B \textbf{76} (2007) 014503.  


\bibitem{Klemm-Luther-Beasley}
 R. A. Klemm, A. Luther, and M. R. Beasley: 
 Phys. Rev. B \textbf{12} (1975) 877. 

\bibitem{Samokhin}
 K. V. Samokhin: 
 Phys. Rev. B \textbf{70} (2004) 104521. 


\bibitem{Takada} 
 S. Takada: 
 Prog. Theor. Phys. \textbf{43} (1970) 27. 

\bibitem{Agterberg-Yang}
 D. F. Agterberg and K. Yang: 
 J. Phys.: Condens. Matter \textbf{13} (2001) 9259. 

\bibitem{Adachi-Ikeda} 
 H. Adachi and R. Ikeda: 
 Phys. Rev. B \textbf{68} (2003) 184510. 

\bibitem{Houzet-Mineev} 
 M. Houzet and V. P. Mineev: 
 Phys. Rev. B \textbf{74} (2006) 144522. 

\bibitem{Yanase-disorder} 
 Y. Yanase: 
 New J. Phys. \textbf{11} (2009) 055056. 


\bibitem{Burkhardt-Rainer} 
 H. Burkhardt and D. Rainer: 
 Ann. Phys. (Leipzig) \textbf{3} (1994) 181. 

\bibitem{Shimahara-FFLO_direction-Q2D} 
 H. Shimahara: 
 Phys. Rev. B \textbf{50} (1994) 12760. 

\bibitem{Shimahara-FFLO_direction-kappa-ET} 
 H. Shimahara: 
 J. Phys. Soc. Jpn. \textbf{66} (1997) 541. 

\bibitem{Buzdin-Kachkachi} 
 A. I. Buzdin and H. Kachkachi: 
 Phys. Lett. A \textbf{225} (1997) 341. 

\bibitem{Vorontsov-Sauls-Graf} 
 A. B. Vorontsov, J. A. Sauls, and M. J. Granf: 
 Phys. Rev. B \textbf{72} (2005) 184501. 

\bibitem{Suginishi-Shimahara-BETS} 
 Y. Suginishi and H. Shimahara: 
 Phys. Rev. B \textbf{74} (2006) 024518. 

\bibitem{Vorontsov-Graf} 
 A. B. Vorontsof and M. J. Graf: 
 Phys. Rev. B \textbf{74} (2006) 172504. 

\bibitem{Shimahara-Moriwake} 
 H. Shimahara and K. Moriwake: 
 J. Phys. Soc. Jpn. \textbf{71} (2002) 1234. 

\bibitem{Kyker-Pickett} 
 A. B. Kyker, W. E. Pickett, and F. Gygi: 
 Phys. Rev. B \textbf{71} (2005) 224517. 



\bibitem{Matsuo-Shimahara-Nagai} 
 S. Matsuo, H. Shimahara, and K. Nagai: 
 J. Phys. Soc. Jpn. \textbf{63} (1994) 2499. 

\bibitem{Shimahara2} 
 H. Shimahara: Phys. Rev. B {\bf 62} (2000) 3524.

\bibitem{Roux-White-Capponi-Poilblanc}
 G. Roux, S. R. White, S. Capponi, and D. Poilblanc: 
 Phys. Rev. Lett. \textbf{97} (2006) 087207. 

\bibitem{Yanase-JPSJ-77-063705}
 Y. Yanase: 
 J. Phys. Soc. Jpn. \textbf{77} (2008) 063705. 

\bibitem{Yokoyama-Onari-Tanaka-FFLO}
 T. Yokoyama, S. Onari, and Y. Tanaka: 
 J. Phys. Soc. Jpn. \textbf{77} (2008) 064711. 



\bibitem{Vorontsov-Sauls-PRB-72-184501}
 A. B. Vorontsov, J. A. Sauls, and M. J. Graf: 
 Phys. Rev. B \textbf{72} (2005) 184501. 

\bibitem{Cui-Hu-PRB-73-214514}
 Q. Cui, C. R. Hu, J. Y. T. Wei, and K. Yang: 
 Phys. Rev. B \textbf{73} (2006) 214514.  

\bibitem{Tanaka-Asano-PRL-98-077001}
 Y. Tanaka, Y. Asano, M. Ichioka, and S. Kashiwaya:  
 Phys. Rev. Lett. \textbf{98} (2007) 077001. 



\bibitem{Radovan-Fortune-et-al}
 H. A. Radovan, N. A. Fortune, T. P. Murphy, S. T. Hannahs, E. C. Palm, 
 S. W. Tozer, and D. Hall: 
 Nature \textbf{425} (2003) 51. 

\bibitem{Bianchi-Movshovich-Capan-et-al}
 A. Bianchi, R. Movshovich, C. Capan, P. G. Pagliuso, and J. L. Sarrao: 
 Phys. Rev. Lett. \textbf{91} (2003) 187004. 

\bibitem{Watanabe-Kasahara-et-al}
 T. Watanabe, Y. Kasahara, K. Izawa, T. Sakakibara, and Y. Matsuda: 
 Phys. Rev. B \textbf{70} (2004) 020506.

\bibitem{Watanabe-Izawa-et-al}
 T. Watanabe, K. Izawa, Y. Kasahara, Y. Haga, Y. Onuki, P. Thalmeier, 
 K. Maki, and Y. Matsuda: 
 Phys. Rev. B \textbf{70} (2004) 184502.

\bibitem{Cpan-Bianchi-et-al}
 C. Capan, A. Bianchi, R. Movshovich, A. D. Christianson, 
 A. Malinowski,	M. F. Hundley, A. Lacerda, P. G. Pagliuso, and 
 J. L. Sarrao: 
 Phys. Rev. B \textbf{70} (2004) 134513. 

\bibitem{Correa-Murphy-et-al}
 V. F. Correa, T. P. Murphy, C. Martin, K. M. Purcell, E. C. Palm, 
 G. M. Schmiedeshoff, J. C. Cooley, and S. W. Tozer: 
 Phys. Rev. Lett. \textbf{98} (2007) 087001. 

\bibitem{Kakuyanagi-Sitoh-et-al}
 K. Kakuyanagi, M. Saitoh, K. Kumagai, S. Takashima, M. Nohara, 
 H. Takagi, and Y. Matsuda: 
 Phys. Rev. Lett. \textbf{94} (2005) 047602. 

\bibitem{Kumagai-Saitoh-et-al}
 K. Kumagai, M. Saitoh, T. Oyaizu, Y. Furukawa, S. Takashima, 
 M. Nohara, H. Takagi, and Y. Matsuda: 
 Phys. Rev. Lett. \textbf{97} (2006) 227002. 

\bibitem{Miclea-Nicklas-et-al}
 C. F. Miclea, M. Nicklas, D. Parker, K. Maki, J. L. Sarrao, 
 J. D. Thompson, G. Sparn, and F. Steglich: 
 Phys. Rev. Lett. \textbf{96} (2006) 117001. 

\bibitem{Gratens-Ferreira-et-al}
 X. Gratens, L. M. Ferreira, Y. Kopelevich, N. F. Oliveira Jr., 
 P. G. Pagliuso, R. Movshovich, R. R. Urbano, J. L. Sarrao, and 
 J. D. Thompson: 
 cond-mat/0608722.

\bibitem{Mitrovic'-Horvatic'-et-al}
 V. F. Mitrovi${\rm \acute {c}}$, M. Horvati${\rm \acute {c}}$, 
 C. Berthier, G. Knebel, G. Lapertot, and J. Flouquet: 
 Phys. Rev. Lett. \textbf{97} (2006) 117002. 


\bibitem{Movshovich-Jaime-et-al}
 R. Movshovich, M. Jaime, J. D. Thompson, C. Petrovic, Z. Fisk, 
 P. G. Pagliuso, and J. L. Sarrao: 
 Phys. Rev. Lett. \textbf{86} (2001) 5152.  

\bibitem{Hall-Palm-et-al}
 D. Hall, E. C. Palm, T. P. Murphy, S. W. Tozer, Z. Fisk, U. Alver, 
 R. G. Goodrich, J. L. Sarrao, P. G. Pagliuso, and T. Ebihara: 
 Phys. Rev. B \textbf{64} (2001) 212508.  

\bibitem{Bianchi-Movshovich-Oeschler-et-al}
 A. Bianchi, R. Movshovich, N. Oeschler, P. Gegenwart, F. Steglich, 
 J. D. Thompson, P. G. Pagliuso, and J. L. Sarrao: 
 Phys. Rev. Lett. \textbf{89} (2002) 137002.  

\bibitem{Izawa-Yamaguchi-et-al}
 K. Izawa, H. Yamaguchi, Y. Matsuda, H. Shishido, R. Settai, 
 and Y. Onuki:  
 Phys. Rev. Lett. \textbf{87} (2001) 057002.  

\bibitem{Aoki-Sakakibara-et-al}
 H. Aoki, T. Sakakibara, H. Shishido, R. Settai, 
 Y. ${\rm \bar{O}}$nuki, P. Miranovi${\rm \acute {c}}$, and K. Machida: 
 J. Phys.: Condens. Matter \textbf{16} (2004) L13. 

\bibitem{Vorontsov-Vekhter}
 A. Vorontsov and I. Vekhter: 
 Phys. Rev. Lett. \textbf{96} (2006) 237001.  


\bibitem{Martin-Agosta-et-al}
 C. Martin, C. C. Agosta, S. W. Tozer, H. A. Radovan, E. C. Palm, 
 T. P. Murphy, and J. L. Sarrao: 
 Phys. Rev. B \textbf{71} (2005) 020503.  


\bibitem{Settai-Shishido-et-al}
 R. Settai, H. Shishido, S. Ikeda, Y. Murakawa, M. Nakashima, D. Aoki, 
 Y. Haga, H. Harima, and Y. ${\rm \bar{O}}$nuki: 
 J. Phys.: Condens. Matter \textbf{13} (2001) L627. 

\bibitem{McCollam-Julian-et-al}
 A. McCollam, S. R. Julian, P. M. C. Rourke, D. Aoki, and J. Flouquet: 
 Phys. Rev. Lett. \textbf{94} (2005) 186401.  

\bibitem{Young-Urbano-et-al}
 B.-L. Young, R. R. Urbano, N. J. Curro, J. D. Thompson, J. L. Sarrao, 
 A. B. Vorontsov, and M. J. Graf: 
 Phys. Rev. Lett. \textbf{98} (2007) 036402.  



\bibitem{Tanatar-Ishiguro-et-al} 
 M. A. Tanatar, T. Ishiguro, H. Tanaka, and H. Kobayashi: 
 Phys. Rev. B \textbf{66} (2002) 134503.  

\bibitem{Uji-Shinagawa-et-al} 
 S. Uji, H. Shinagawa, T. Terashima, T. Yakabe, Y. Terai, 
 M. Tokumoto, A. Kobayashi, H. Tanaka, and H. Kobayashi: 
 Nature \textbf{410} (2001) 908.  

\bibitem{Balicas-Brooks-et-al} 
 L. Balicas, J. S. Brooks, K. Storr, S. Uji, M. Tokumoto, H. Tanaka, 
 H. Kobayashi, A. Kobayashi, V. Barzykin, and L. P. Gor'kov: 
 Phys. Rev. Lett. \textbf{87} (2001) 067002.  

\bibitem{Uji-Terashima-et-al} 
 S. Uji, T. Terashima, M. Nishimura, Y. Takahide, T. Konoike, 
 K. Enomoto, H. Cui, H. Kobayashi, A. Kobayashi, H. Tanaka, 
 M. Tokumoto, E. S. Choi, T. Tokumoto, D. Graf, and J. S. Brooks: 
 Phys. Rev. Lett. \textbf{97} (2006) 157001.  



\bibitem{Manalo-Klein}
 S. Manalo and U. Klein: 
 J. Phys.: Condens. Matter \textbf{12} (2000) L471. 

\bibitem{Singleton-Symington-et-al}
 J. Singleton, J. A. Symington, M.-S. Nam, A. Ardavan, M. Kurmoo, 
 and P. Day: 
 J. Phys.: Condens. Matter \textbf{12} (2000) L641. 

\bibitem{Lortz-Wang-et-al}
 R. Lortz, Y. Wang, A. Demuer, P. H. M. B${\rm \ddot{o}}$ttger, 
 B. Bergk, G. Zwicknagl, Y. Nakazawa, and J. Wosnitza: 
 Phys. Rev. Lett. \textbf{99} (2007) 187002. 



\bibitem{Zwierlein-Schirotzek-et-al}
 M. W. Zwierlein, A. Schirotzek, C. H. Schunck, and W. Ketterle: 
 Science \textbf{311} (2006) 492. 

\bibitem{Partridge-Li-et-al}
 G. B. Partridge, W. Li, R. I. Kamar, Y. Liao, and R. G. Hulet: 
 Science \textbf{311} (2006) 503. 

\bibitem{Anderson-Brinkman}
 P. W. Anderson and W. F. Brinkman: 
 Phys. Rev. Lett. \textbf{30} (1973) 1108.  

\bibitem{Nakajima-RPA}
 S. Nakajima: 
 Prog. Theor. Phys. \textbf{50} (1973) 1101. 

\bibitem{Miyake-Schmitt-Varma-RPA}
 K. Miyake, S. Schmitt-Rink, and C. M. Varma: 
 Phys. Rev. B \textbf{34} (1986) 6554. 

\bibitem{Scalapino-Loh-Hirsch-RPA}
 D. J. Scalapino, E. Loh, Jr., and J. E. Hirsch: 
 Phys. Rev. B \textbf{34} (1986) 8190. 

\bibitem{Shimahara-Takada-RPA}
 H. Shimahara and S. Takada: 
 J. Phys. Soc. Jpn. \textbf{57} (1988) 1044. 


\bibitem{Yoshino-Shodai-SM-133-55}
 H. Yoshino, S. Shodai, and K. Murata: 
 Synth. Met. \textbf{133} (2003) 55. 

\bibitem{Lebed-Ha-PRB-71-132504}
 A. G. Lebed, Heon-Ick Ha, and M. J. Naughton: 
 Phys. Rev. B \textbf{71} (2005) 132504. 

\bibitem{Shimahara-PRB-61-R14938}
 H. Shimahara: 
 Phys. Rev. B \textbf{61} (2000) R14938. 


\bibitem{Berezinskii-JETP-Lett-20-287}
 V. L. Berezinskii: JETP Lett. \textbf{20} (1974) 287. 

\bibitem{Balatsky-Abrahams-PRB-45-13125}
 A. Balatsky and E. Abrahams: Phys. Rev. B \textbf{45} (1992) 13125.

\bibitem{Bergeret-Volkov-RMP-77-1321}
 F. S. Bergeret, A. F. Volkov, and K. B. Efetov: 
 Rev. Mod. Phys. \textbf{77} (2005) 1321. 


\bibitem{Shigeta-Onari}
 K. Shigeta, S. Onari, K. Yada, and Y. Tanaka:  
 Phys. Rev. B \textbf{79} (2009) 174507. 



\bibitem{Tanaka-Golubov-PRL-98-037003}
 Y. Tanaka and A. A. Golubov: 
 Phys. Rev. Lett. \textbf{98} (2007) 037003. 

\bibitem{Tanaka-Golubov-PRL-99-037005}
 Y. Tanaka, A. A. Golubov, S. Kashiwaya, and M. Ueda: 
 Phys. Rev. Lett. \textbf{99} (2007) 037005. 

\bibitem{Tanaka-Tanuma-PRB-76-054533}
 Y. Tanaka, Y. Tanuma, and A. A. Golubov: 
 Phys. Rev. B \textbf{76} (2007) 054522. 

\bibitem{Tanaka-Asano-PRB-77-220504R}
 Y. Tanaka, Y. Asano, and A. A. Golubov: 
 Phys. Rev. B \textbf{77} (2008) 220504(R). 

\bibitem{Yokoyama-Tanaka-PRB-78-012508}
 T. Yokoyama, Y. Tanaka, and A. A. Golubov: 
 Phys. Rev. B \textbf{78} (2008) 012508. 

\bibitem{Tanuma-Hayashi-PRL-102-11703}
 Y. Tanuma, N. Hayashi, Y. Tanaka, and A. A. Golubov: 
 Phys. Rev. Lett. \textbf{102} (2009) 117003.


\end{thebibliography}
\end{document}